\documentclass[letter]{mn2e}
\usepackage{psfig}
\newcommand{\HI}{{\sc H\,i}}

\newcommand{\kms}{$\,$km$\,$s$^{-1}$}
\newcommand{\ltsima} {$\; \buildrel < \over \sim \;$}
\newcommand{\gtsima} {$\; \buildrel > \over \sim \;$}
\newcommand{\lta} {\lower.5ex\hbox{\ltsima}}
\newcommand{\gta} {\lower.5ex\hbox{\gtsima}}
\newcommand{\sauron}{{\texttt {SAURON}}}
\newcommand{\Oiii}{[{\sc O$\,$iii}]}

\title [The dark matter halo in NGC~2974]
{The shape of the dark matter halo in the early-type galaxy NGC~2974}

\author[A. Weijmans et al.]
   {Anne-Marie Weijmans$^{1}$\thanks{E-mail: weijmans@strw.leidenuniv.nl},  Davor Krajnovi\'c$^{2}$, Glenn van de Ven$^{3}$\thanks{Hubble Fellow}, \and
  Tom~A. Oosterloo$^{4,5}$, Raffaella Morganti$^{4,5}$ and  P.~T. de Zeeuw$^{1,6}$  \\  
$^1$ Sterrewacht Leiden, Leiden University,  Postbus 9513, 2300 RA Leiden, The Netherlands  \\
$^2$ Denys Wilkinson Building, University of Oxford, Keble Road, Oxford, UK \\
$^3$ Institute for Advanced Study, Einstein Drive, Princeton, NJ 08540, USA \\
$^4$ Netherlands Foundation for Research in Astronomy, Postbus 2,
      7990 AA Dwingeloo, The Netherlands \\
$^5$ Kapteyn Astronomical Institute, University of Groningen,  Postbus 800, 9700 AV Groningen, The Netherlands \\
$^6$ European Southern Observatory, Karl-Schwarzschild Strasse 2, 85748 Garching bei Munchen, Germany
   }

\pagerange{\pageref{firstpage}--\pageref{lastpage}} \pubyear{2007}

\begin{document}

\maketitle

\label{firstpage}

\begin{abstract} 
  We present \HI\ observations of the elliptical galaxy NGC 2974,
  obtained with the Very Large Array. These observations reveal that
  the previously detected \HI\ disc in this galaxy (Kim et al.\ 
  1988\nocite{1988ApJ...330..684K}) is in fact a ring. By studying the
  harmonic expansion of the velocity field along the ring, we
  constrain the elongation of the halo and find that the underlying
  gravitational potential is consistent with an axisymmetric shape.
  
  We construct mass models of NGC 2974 by combining the \HI\ rotation
  curve with the central kinematics of the ionised gas, obtained with
  the integral-field spectrograph \sauron. We introduce a new way of
  correcting the observed velocities of the ionised gas for asymmetric
  drift, and hereby disentangle the random motions of the gas caused
  by gravitational interaction from those caused by turbulence. To
  reproduce the observed flat rotation curve of the \HI\ gas, we need
  to include a dark halo in our mass models. A pseudo-isothermal
  sphere provides the best model to fit our data, but we also tested
  an NFW halo and Modified Newtonian Dynamics (MOND), which fit the
  data marginally worse.
  
  The mass-to-light ratio $M/L_I$ increases in NGC~2974 from 4.3
  $M_\odot/L_{\odot,I}$ at one effective radius to 8.5
  $M_\odot/L_{\odot,I}$ at 5 $R_e$. This increase of $M/L$ already
  suggests the presence of dark matter: we find that within 5 $R_e$
  at least 55 per cent of the total mass is dark.
  \\

    \date{}
\end{abstract}



\section{Introduction}
\label{sec:introduction}

Although the presence of dark matter dominated haloes around spiral
galaxies is well established (e.g. van Albada et al.
1985\nocite{1985ApJ...295..305V}), there is still some controversy
about their presence around early-type galaxies.  Spiral galaxies
often contain large regular \HI\ discs, which allow us to obtain
rotation curves out to large radii, and therefore we can constrain the
properties of their dark haloes. But these discs are much rarer in
elliptical galaxies (e.g. Bregman, Hogg \& Roberts
1992\nocite{1992ApJ...387..484B}), so that for this class of galaxies
we are often required to use other tracers to obtain velocity
measurements, such as stellar kinematics, planetary nebulae or
globular clusters. These tracers however are not available for all
early-type galaxies, and give mixed results (e.g.  Rix et al.
1997\nocite{1997ApJ...488..702R}, Romanowsky et al.
2003\nocite{2003Sci...301.1696R}, Bridges et al.
2006\nocite{2006MNRAS.373..157B}).

With the increase in sensitivity of radio telescopes, it has been
discovered that many early-type galaxies in the field do contain \HI\ 
gas, though with smaller surface densities than in spiral galaxies
(e.g. Morganti et al. 2006\nocite{2006MNRAS.371..157M}). The average
\HI\ surface density in the Morganti et al. sample is around $1
M_\odot$ pc$^{-2}$, which is far below the typical value for spiral
galaxies (4 - 8$ M_\odot$ pc$^{-2}$, e.g. Cayatte et al.
1994\nocite{1994AJ....107.1003C}). This would explain why previously
only the most gas-rich early-type galaxies were detected in \HI.
Morganti et al. find that \HI\ can be present in different
morphologies: \HI\ discs seem to be as common as off-set clouds and
tails, though they occur mostly in the relatively gas-rich systems.

Recently rotation curves of \HI\ discs in low surface brightness
galaxies and dwarf galaxies, complemented with H$\alpha$ observations,
have been used not only to confirm the existence of dark matter
haloes, but also to obtain estimates on the inner slope of the
density profiles of the haloes (e.g. van den Bosch et al.
2000\nocite{2000AJ....119.1579V}; Weldrake, de Blok \& Walter
2003\nocite{2003MNRAS.340...12W}). Simulations within a cold dark
matter (CDM) cosmology yield haloes with cusps in their centres (NFW
profiles, see Navarro, Frenk \& White
1996\nocite{1996ApJ...462..563N}), but observations suggest
core-dominated profiles (e.g. de Blok \& Bosma
2002\nocite{2002A&A...385..816D}; de Blok
2005\nocite{2005ApJ...634..227D}).

Detailed studies of rotation curves of early-type galaxies that
contain \HI\ discs are sparser, due to lack of spatial resolution: to
detect low \HI\ surface densities, larger beams are needed.  Also,
only few early-type galaxies have \HI\ discs that are extended and
regular enough to allow for detailed studies.  Comparing $M/L$ values
at large radii, derived from \HI\ velocities, to $M/L$ at smaller
radii measured from ionised gas kinematics, the conclusion is that
early-type galaxies also have dark matter dominated haloes (e.g.
Bertola et al.  1993\nocite{1993ApJ...416L..45B}; Morganti et al.
1997\nocite{ 1997AJ....113..937M}; Sadler et al.
2000\nocite{2000AJ....119.1180S}).  Franx, van Gorkom \& de Zeeuw
(1994)\nocite{1994ApJ...436..642F} used the \HI\ ring of the
elliptical galaxy IC~2006 to determine not only the mass, but also the
shape of the dark halo. They concluded that IC~2006 is surrounded by
an axisymmetric dark halo, using the geometry of the ring and an
harmonic expansion of its velocity map.

In this paper, we present a similar analysis of the regularly rotating
\HI\ ring around the elliptical (E4) field galaxy NGC~2974. Kim et al.
(1988)\nocite{1988ApJ...330..684K} observed this galaxy before in
\HI\, but their data had lower spatial resolution than ours, and they
found a filled disc instead of a ring. Cinzano \& van der Marel
(1994)\nocite{{1994MNRAS.270..325C}} found an embedded stellar disc in
their dynamical model of this galaxy, based upon long-slit
spectroscopic data, but Emsellem, Goudfrooij \& Ferruit
(2003)\nocite{2003MNRAS.345.1297E} constructed a dynamical model of
NGC~2974 based on TIGER integral-field spectrography and long-slit
stellar kinematics, that does not require a hidden disc structure.
They did report the detection of a two-arm gaseous spiral in the inner
200 pc of NGC~2974 from high resolution WFPC2 imaging.  Krajnovi\'c et
al.  (2005)\nocite{2005MNRAS.357.1113K} constructed axisymmetric
dynamical models of both the stars and ionised gas based upon \sauron\ 
integral-field data. These models require a component with high
angular momentum, consisting of a somewhat flattened distribution of
stars, though not a thin stellar disc.  Emsellem et al.
(2007)\nocite{2007MNRAS.379..401E} classify NGC~2974 as a fast rotator,
which means that it possesses large-scale rotation and that its
angular momentum is well defined. Some of the characteristics of
NGC~2974 are given in Table~\ref{tab:ngc2974}.

For our analysis of NGC~2974 we combine kinematics of neutral gas,
obtained from our observations with the Very Large Array (VLA), with
that of ionised gas, obtained with the integral-field spectrograph
\sauron\ (Bacon et al.  2001\nocite{2001MNRAS.326...23B}). This
combination of a small scale two-dimensional gas velocity map in the
centre of the galaxy, and a \HI\ velocity map at the outskirts, allows
measurements of a rotation curve ranging from 100 pc within the centre
of the galaxy to 10 kpc at the edges of the \HI\ ring. We use this
rotation curve, together with ground- and space based optical imaging,
to determine the dark matter content in NGC~2974, and to constrain the
shape of the dark halo.

In section 2, we discuss the two datasets and their reduction, and
describe the \HI\ ring. We concentrate on the analysis of the velocity
maps in section 3. Section 4 is devoted to the rotation curve that we
extract from the velocity maps, and in section 5 we show mass models
with various halo models, and find the best fit to the rotation curve.
Section 6 summarizes our results.

\begin{table}
\label{tab:ngc2974}
\begin{center}
\begin{tabular}{l|r}
\hline\hline
Parameter & Value \\
\hline
Morphological Type & E4 \\
$M_B$ (mag) & -20.07 \\
Effective $B-V$ (mag) & 0.93 \\
PA ($^{\circ}$) & 41 \\
Distance modulus (mag) & 31.60 \\
Distance (Mpc) & 20.89 \\
Distance scale (pc arcsec$^{-1}$) & 101.3 \\ 
Effective radius & 24$^{\prime \prime}$ \\
\hline
\end{tabular}
\end{center}
\caption{Properties of NGC~2974. The values are taken from the Lyon/Meudon Extragalactic Database (LEDA) and corrected for the distance modulus, which is taken from the surface brightness fluctuation measurements by Tonry et al. (2001)
. Note that 0.06 mag is subtracted to adjust to the Cepheid zeropoint of Freedman et al. (2001)
; see Mei et al. (2005), 
 section 3.3, for a discussion. The effective radius is taken from Cappellari et al. (2006).}
\end{table}


\section{Observations and data reduction}
\label{sec:observations}

\subsection{VLA observations}

\begin{figure}
\centerline{\psfig{figure=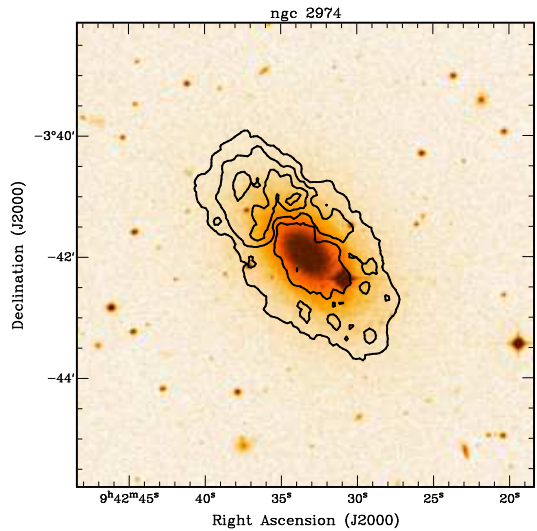,angle=0,width=7.0cm}}
\caption{Total \HI\ intensity contours superimposed onto the Digital Sky Survey optical image of NGC~2974. Contour levels are 1, 3, 5 and 7$\times 10^{20}$ cm$^{-2}$. The beamsize is 19.9 $\times$ 17.0 arcseconds.}
\label{fig:optHI}
\end{figure}

Earlier VLA observations (Kim et al. 1988\nocite{1988ApJ...330..684K})
of NGC~2974 showed that this galaxy contains a significant amount of
\HI\ that, in their observations, appears to be distributed in a
regularly rotating disc.  Given the modest spatial and velocity
resolution of those observations, we re-observed NGC 2974 with the VLA
C-array while also using a different frequency setup that allows us to
study this galaxy at both higher spatial and higher velocity
resolution. The observations were performed on 11 and 19 September
2005 with a total on-source integration time of 15 hours. In each
observation, two partially overlapping bands of 3.15 MHz and 64
channels were used. The two bands were offset by 500 \kms\ in central
velocity. This frequency setup allows us to obtain good velocity
resolution over a wide range of velocities (about 1080 \kms).

The data were calibrated following standard procedures using the
MIRIAD software package (Sault, Teuben \& Wright 1995\nocite{xxx}).  A
spectral-line data cube was made using robust weighting (robustness =
1.0) giving a spatial resolution of $19.9^{\prime\prime}\times
17.0^{\prime\prime}$ and a velocity resolution of 20.0 \kms\ (after
Hanning smoothing).  The noise in the final datacube is 0.23 mJy
beam$^{-1}$.

\begin{figure*}
\centerline{\psfig{figure=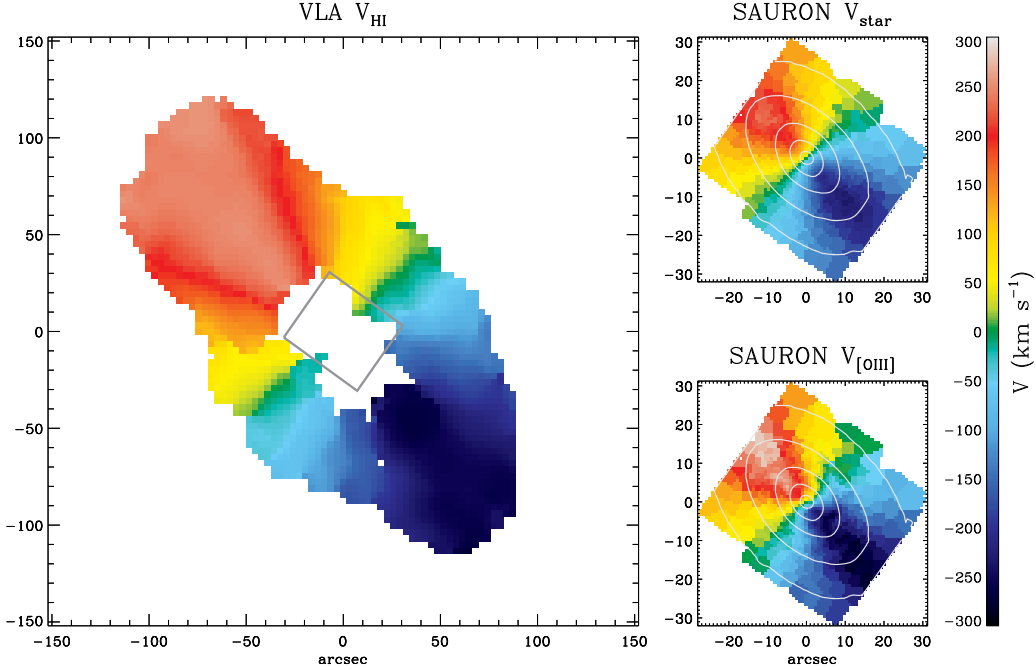,angle=0,width=14.0cm}}
\caption{Velocity maps of the neutral hydrogen (VLA) and ionised gas and stars (\sauron) in NGC~2974. Both the stars and the neutral and ionised gas are well aligned. The maps are orientated so that North is up and East is to the left. The grey box in the VLA map encloses the \sauron\ fields shown at the right.}
\label{fig:velmap}
\end{figure*}

To construct the total \HI\ image, a mask was created using a datacube
that was smoothed to about twice the spatial resolution and that was
clipped at twice the noise of that smoothed datacube. The resulting
total \HI\ is shown in Figure~\ref{fig:optHI}, and our observations
show that the \HI\ is distributed in a regular rotating ring instead
of a filled disc. The inner radius of the ring is approximately
$50^{\prime \prime}$ ($\sim$ 5 kpc) and extends to $120^{\prime
  \prime}$, which corresponds to 12 kpc, or 5 effective radii (1 $R_e$
= 24$^{\prime \prime}$).

The \HI\ velocity field was derived by fitting Gaussians to the
spectra at those positions where signal is detected in the total \HI\ 
image. The resulting velocity map is shown in Figure~\ref{fig:velmap}.
Typical errors on this map are 5 - 10 \kms.

We find a total mass of $5.5 \times 10^8 M_\odot$ for the \HI\ gas
content of the ring, which is in agreement with Kim et al.
(1988)\nocite{1988ApJ...330..684K}, if we correct for the difference
in assumed distance modulus. The amount and morphology of the \HI\ observed in
NGC~2974 are not unusual for early-type galaxies. Oosterloo et al.
(2007)\nocite{2007A&A...465..787O} have found that between 5 and 10 per cent
of early-type galaxies show \HI\ masses well above $10^9 M_\odot$,
while the fraction of detections increases further for lower \HI\ 
masses (Morganti et al. 2006\nocite{2006MNRAS.371..157M}). The
majority of the \HI-rich systems have the neutral hydrogen distributed
in disc/ring like structures (often warped) with low surface
brightness density and no or little ongoing star formation, as
observed in NGC~2974. However, there is a region in the
North-East of the \HI\ ring where the surface density is higher, and the
gas could be forming stars.  Jeong et al.
(2007)\nocite{2007MNRAS.376.1021J} published UV imaging of NGC~2974,
obtained with GALEX. Their images reveal indeed a region of increased
starformation in the North-East of the galaxy, as well as a
starforming ring at the inner edges of the \HI\ ring.

At least some of the most \HI\ rich structures are the results of
major mergers (see e.g. Serra et al.
2006\nocite{2006A&A...453..493S}). For the systems with less extreme
\HI\ masses, like NGC~2974, the origin of the gas is less clear.
Accretion of small companions is a possibility, but smooth, cold
accretion from the intergalactic medium (IGM) is an alternative
scenario.

\subsection{SAURON observations}

Maps of the stellar and ionised gas kinematics of NGC~2974, obtained
with the integral-field spectrograph \sauron\, were presented in
Emsellem et al. (2004)\nocite{2004MNRAS.352..721E} and Sarzi et
al. (2006)\nocite{2006MNRAS.366.1151S}, respectively, and we refer the
reader to these papers for the methods of data reduction and
extraction of the kinematics. 

In Figure~\ref{fig:velmap} we compare both the \sauron\ velocity maps
of stars and \Oiii\ with the velocity map of the \HI\ ring. Stars and
gas are well aligned, and the transition between the ionised and the
neutral gas seems to be smooth, suggesting that they form one single
disc. The twist in the velocity map of the ionised gas in the inner
$4^{\prime\prime}$ is likely caused by the inner bar of this galaxy
(Emsellem et al. 2003\nocite{2003MNRAS.345.1297E},
Krajnovi\'c et al. 2005\nocite{2005MNRAS.357.1113K}).


\section{Analysis of velocity fields}

We used kinemetry (Krajnovi\'c et al.
2006\nocite{2006MNRAS.366..787K}) to analyse the \sauron\ and VLA
velocity maps. In our application to a gas disc, kinemetry reduces to
the tilted-ring method (Begeman 1978\nocite{1987PhDT.......199B}). The
velocity along each elliptical ring is expanded in Fourier components
(e.g. Franx et al.  1994\nocite{1994ApJ...436..642F}; Schoenmakers,
Franx \& de Zeeuw 1997\nocite{1997MNRAS.292..349S}):

\begin{equation}
\label{eq:expan}
V_{\mathrm{los}}(R,\phi) = V_{\mathrm{sys}}(R) + \sum_{n=1}^{N} c_n(R) 
\cos n\phi + s_n(R) \sin n\phi, 
\end{equation}

\noindent
where $V_{\mathrm{los}}$ is the observed velocity, $R$ is the length
of the semimajor axis of the elliptical ring, $\phi$ the azimuthal
angle, measured from the projected major axis of the galaxy,
$V_{\mathrm{sys}}$ the systemic velocity of the ring and $c_n$ and
$s_n$ are the coefficients of the harmonic expansion. The $c_1$ term
relates to the circular velocity $V_c$ in the disc, so that $c_1 = V_c
\sin i$, where $i$ is the inclination of the gas disc.  Assuming that
motions in the ring are intrinsically circular and that the ring is
infinitely thin, the inclination can be inferred from the flattening
$q$ of the fitted ellipse: $\cos i = q$.

If a gas disc only displays pure circular motions, all harmonic terms
other than $c_1$ in Equation~(\ref{eq:expan}) are zero.  Noncircular
motions, originating from e.g. inflows caused by spiral arms or bars,
or a triaxial potential, will cause these terms to deviate from zero.
Alternatively, also wrong input parameters of the ring (which are
flattening $q$, position angle $\Gamma$ and the coordinates of the
centre of the ellipse) will result in specific patterns in these
terms, see e.g. van der Kruit \& Allen
(1978)\nocite{1978ARA&A..16..103V}, Schoenmakers et al.
(1997)\nocite{1997MNRAS.292..349S} and also Krajnovi\'c et al.
(2006)\nocite{2006MNRAS.366..787K} for details.  Therefore, the
flattening and position angle of each ring are determined by
minimising $s_1, s_3$ and $c_3$ along that ring. The centre is kept
constant and is chosen to coincide with the position of maximal flux
in the galaxy.

\subsection{Noncircular motions}

\begin{figure}
\centerline{\psfig{figure=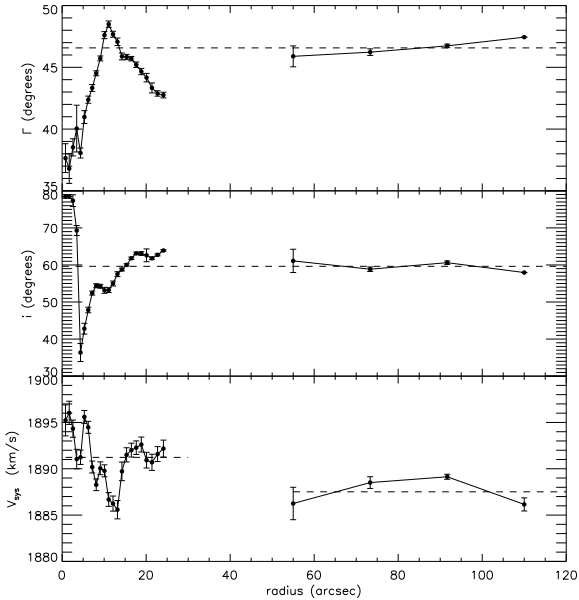,angle=0,width=8cm}}
\caption{From top to bottom: position angle $\Gamma$, inclination $i$ and systemic velocity of the rings that are fitted to the \sauron\ and VLA velocity fields of NGC~2974. The position angle indicates the receding side of the galaxy and is measured North through East. The dashed lines in the top two panels indicate the mean value of $\Gamma$ and $i$ generated by the \HI\ rings. The bottom panel has two dashed lines, indicating the mean systemic velocity of the \sauron\ and VLA rings separately.}
\label{fig:kinthree}
\end{figure}

\begin{figure}
  \centerline{\psfig{figure=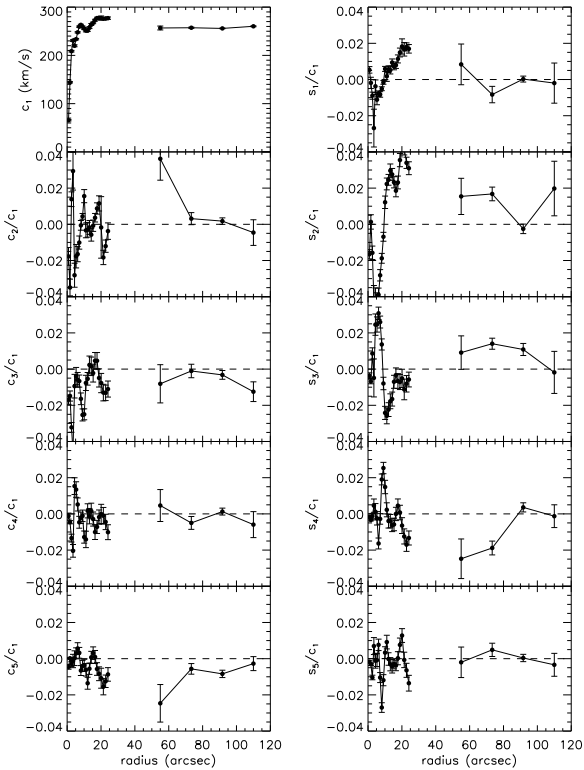,angle=0,width=8cm}}
\caption{Coefficients of the harmonic expansion on the \sauron\ and VLA velocity fields. All except $c_1$ are normalized with respect to $c_1$.}
\label{fig:kinhigh}
\end{figure}

Figure~\ref{fig:kinthree} shows the properties of the elliptic rings
that were fitted to the \sauron\ and VLA velocity fields, and
Figure~\ref{fig:kinhigh} shows the resulting harmonic terms. The
datapoints of the VLA data are separated by approximately one
beamsize. Error bars were calculated by constructing 100 Monte Carlo
realisations of the velocity fields, where the measurement errors of
the maps were taken into account. 

Both the position angles and the inclinations of the rings show some
variation in the \sauron\ field, but are very stable in the VLA field.
The dashed line in the top two panels of Figure~\ref{fig:kinthree}
indicates the mean value of the position angle and inclination of the
\HI\ data, which are $\Gamma = 47 \pm 1^\circ$ and $i = 60 \pm
2^\circ$. Here, $\Gamma$ is the position angle of the receding side of
the galaxy, measured North through East. The systemic velocities
(lower panel of Figure~\ref{fig:kinthree}) have been corrected for
barycentric motion and are in good agreement. For the \sauron\ field
we find a systemic velocity of $1891 \pm 3$ \kms, while for the VLA
field we find $1888 \pm 2$ \kms.  The dashed lines give both these
mean velocities. Both the inclination and the systemic velocity that
we find are in agreement with previous studies (Cinzano \& Van der
Marel 1994\nocite{1994MNRAS.270..325C}, Emsellem et al.
2003\nocite{2003MNRAS.345.1297E}, Krajnovi\'c et al.
2005\nocite{2005MNRAS.357.1113K}).

The harmonic terms are shown in Figure~\ref{fig:kinhigh}. All terms
are normalised with respect to $c_1$. From $c_1$ we see that the
velocity curve of the gas rises steeply in the centre, but flattens
out at larger radii. This already suggests that a dark halo is present
around this galaxy. In \S\ref{sec:rot} we will analyse the rotation
curve in more detail.

The other terms have small amplitudes, and are small compared to $c_1$
($<$ 4 per cent). We do not observe signatures that could indicate
incorrect ring parameters, as described in Schoenmakers et al.
(1997)\nocite{1997MNRAS.292..349S} and Krajnovi\'c et al.
(2006)\nocite{2006MNRAS.366..787K}.

\subsection{Shape of the gravitational potential}
\label{sec:elong}

Following Schoenmakers et al. (1997)\nocite{1997MNRAS.292..349S}, we
calculate the elongation of the potential from the harmonic terms.
Using epicycle theory these authors showed that an $\cos m\phi$-term
perturbation of the potential results in signal in the $m-1$ and $m+1$
coefficients of the harmonic expansion in Equation~(\ref{eq:expan}).

We assume that the potential of NGC~2974 is affected by an $m=2$
perturbation, which could correspond to a perturbation by a bar. We
assume that the galaxy is not affected by lopsidedness, warps or
spiral arms. To first order, the potential of the galaxy in the plane
of the gas ring can then be written as:

\begin{equation}
\label{eq:pot_bar}
\Phi(R, \phi) = \Phi_0(R) + \Phi_2(R) \cos 2\phi,
\end{equation} 

\noindent
with $\Phi_2(R)\ll\Phi_0(R)$. As explained in Schoenmakers et al.
(1997)\nocite{1997MNRAS.292..349S}, the elongation of the potential
$\epsilon_{\mathrm{pot}}$ in the plane of the gas is in this case
given by:

\begin{equation}
\label{eq:elong}
\epsilon_{\mathrm{pot}}\sin 2\varphi = \frac{(s_3- s_1)}{c_1}\frac{(1+2q^2+5q^4)}{1-q^4},
\end{equation}

\noindent
where $\varphi$ is one of the viewing angles of the galaxy, namely the
angle between the minor axis of the galaxy and the observer, measured
in the plane of the disc. This viewing angle is in general unknown, so
that from this formula only a lower limit on the elongation can be
derived. Schoenmakers (1998)\nocite{1998ASPC..136..240S} used this
method in a statistical way and found an average elongation
$\epsilon_{\mathrm{pot}} = 0.044$ for a sample of 8 spiral galaxies. 

\begin{figure}
    \centerline{\psfig{figure=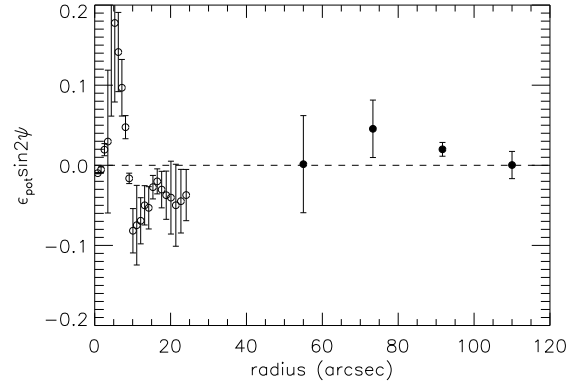,angle=0,width=8cm}}
\caption{Elongation of the potential ($\epsilon_{\mathrm{pot}}\sin 2\varphi$) of NGC~2974 as a function of radius. Open dots denote measurements from the ionised gas, filled dots represent the \HI\ gas. The elongation as measured from the ionised gas is varying, due to the bar and high random motions of the gas. The cold neutral gas yields a more reliable value for the elongation, and shows that the potential is consistent with axisymmetry. }
\label{fig:elong}
\end{figure}

We calculated the elongation at different radii in NGC~2974, and the
result is plotted in Figure~\ref{fig:elong}. As in Schoenmakers et al,
we did not fix $\Gamma$ and $q$ when determining the harmonic terms,
because an offset in $\Gamma$ or $q$ introduces extra signal in $c_1,
s_1$ and $s_3$, that would then be attributed to the elongation of the
potential.

Although the ionised gas has high random motions (see also
\S~\ref{sec:rot}) and therefore the calculated elongation is probably
only approximate, it is striking that the elongation changes sign
around $10^{\prime \prime}$. The potential in the inner 10 arcseconds
has a rather high elongation $\epsilon_{\mathrm{pot}}\sin 2 \varphi =
0.10 \pm 0.08$, while outside this region the elongation as measured
from the ionised gas is $\epsilon_{\mathrm{pot}}\sin 2 \varphi =
-0.047 \pm 0.020$. The change of sign could be the result of the bar
system in NGC~2974, with the direction along which the potential is
elongated changing perpendiculary. It is worth mentioning here that
Krajnovi\'c et al.  (2005)\nocite{2005MNRAS.357.1113K} find a ring in
the \Oiii\ equivalent width map, with a radius of 9$^{\prime \prime}$.
Their data suggest also the presence of a (pseudo-)ring around
28$^{\prime \prime}$, and Jeong et al.
(2007)\nocite{2007MNRAS.376.1021J} find a ring with a radius of $\sim
60^{\prime \prime}$ in their GALEX UV map, which is where our \HI\ 
starts. Assuming that these three rings are resonances of a single
bar, Jeong et al. (2007)\nocite{2007MNRAS.376.1021J} deduce a pattern
speed of $78 \pm 6$ \kms\ kpc$^{-1}$. In addition to the large scale
bar, Emsellem et al. (2003)\nocite{2003MNRAS.345.1297E} postulate a
small nuclear bar ($\sim 3^{\prime \prime}$).

The \HI\ gas is more suitable for measuring the elongation of the
potential, since the cold gas has a small velocity dispersion (typical
values $< 10 $ \kms) and is on nearly circular orbits.  Taking the
mean value of the elongation as obtained from the \HI\ field, we find
$\epsilon_{\mathrm{pot}}\sin 2 \varphi = 0.016 \pm 0.022$.  We
conclude that the potential of NGC~2974 is well approximated by an
axisymmetric one.

\section{Rotation curve}
\label{sec:rot}

To find the rotation curve of NGC~2974, we subtract the systemic
velocities from the ionised and neutral gas velocity fields
separately. Next, we fix $\Gamma = 47^\circ$ and $q=0.50$ (or
equivalently $i=60^\circ$) of the ellipses to the mean values obtained
from the neutral gas, and rerun kinemetry on both the velocity maps,
now forcing the position angle and flattening to be the same
everywhere in the gas disc. Also, because velocity is an odd moment,
the even terms in the harmonic expansion should be zero, and are not
taken into account during the fit (see Krajnovi\'c et al.
2006\nocite{2006MNRAS.366..787K}). The rotation curve of the ionised
gas is shown in Figure~\ref{fig:fit_vphi} (open diamonds).

The ionised gas has a high observed velocity dispersion
$\sigma_{\mathrm{obs}}$, exceeding 250 \kms\ in the centre of the
galaxy. Three phenomena can contribute to the observed velocity
dispersion of a gas: thermal motions, turbulence and gravitational
interactions:

\begin{equation}
\label{eq:cause_disp}
\sigma_{\mathrm{obs}}^2 = \sigma_{\mathrm{thermal}}^2 + \sigma_{\mathrm{turb}}^2 + 
\sigma_{\mathrm{grav}}^2.
\end{equation}

\noindent
The thermal velocity dispersion is always present, and caused by the
thermal energy of the gas molecules:

\begin{equation}
\sigma_{\mathrm{thermal}}^2 = \frac{kT}{m},
\end{equation}

\noindent
where $k$ is Boltzmann's constant, $T$ the temperature of the gas and
$m$ the typical mass of a gas particle. The contribution of
$\sigma_{\mathrm{thermal}}$ to the total velocity distribution in
ionised gas is small: a typical temperature for ionised gas is $10^4$
K, which implies $ \sigma_{\mathrm{thermal}} \sim 10$ km/s. 

Turbulence can be caused by e.g. internal motions within the gas
clouds orshocks induced by a non-axisymmetric perturbation to the
potential, such as a bar. This increases the dispersion, but has a
negligible effect on the circular velocity of the gas. In contrast,
gravitational interactions of individual gas clouds not only increase
random motions of the clouds and therefore their dispersion, but also
lower the observed velocity.  To correct for this last effect, we need
to apply an asymmetric drift correction to recover the true circular
velocity.

Unfortunately, it is not possible {\it a priori} to determine which
fraction of the high velocity dispersion in the ionised gas is caused
by turbulence and which by gravitational interactions. We therefore
now first investigate the effect of asymmetric drift on the rotation
curve of the ionised gas.

\subsection{Asymmetric drift correction}
\label{sec:adcshort}

Due to gravitational interactions of gas clouds on circular orbits,
the observed velocity is lower than the circular velocity connected to
the gravitational potential. Since we are interested in the mass
distribution of NGC~2974, we need to trace the potential, and
therefore we have to increase our observed velocity with an asymmetric
drift correction, to obtain the true circular velocity. We follow the
formalism described in Appendix~\ref{sec:adc}, which is based on the
Jeans equations and the higher order velocity moments of the
collisionless Boltzmann equation.

We assume that the galaxy is axisymmetric, which is a valid approach
given the low elongation of the potential that we derived in
section~\ref{sec:elong}. Further we assume that the gas lies in a thin
disc.

We fit the prescription that Evans \& de Zeeuw
(1994\nocite{1994MNRAS.271..202E}) used for their power-law models to
the rotation curve extracted from the ionised gas,

\begin{equation}
\label{eq:vmod}
v_{\mathrm{mod}} = \frac{V_\infty R}{R_{\mathrm{mod}}},
\end{equation}

\noindent
where $V_\infty$ is the rotation velocity at large radii, and we
introduce

\begin{equation}
\label{eq:rmod}
R_{\mathrm{mod}}^2 = R^2 + R_c^2, 
\end{equation}

\noindent
with $R_c$ the core radius of the model. This is
Equation~(\ref{eq:powerlaw}) evaluated in the plane of the disc ($z =
0$), with a flat rotation curve at large radii ($\beta = 0$). Since we
observe the gas only in the equatorial plane of the galaxy, we cannot
constrain the flattening of the potential $q_\Phi$. We therefore
assumed a spherical potential $q_\Phi = 1$, which is not a bad
approximation even if the density distribution is flattened, since the
dependence on $q_\Phi$ is weak.  Moreover, even though the density
distribution of most galaxies is clearly flattened, the potential is
in general significantly rounder than the density.  For example, an
axisymmetric logarithmic potential is only about a third as flattened
as the corresponding density distribution (e.g. \S2.2.2 of Binney \&
Tremaine 1987\nocite{1987BT}).

\begin{figure}
    \centerline{\psfig{figure=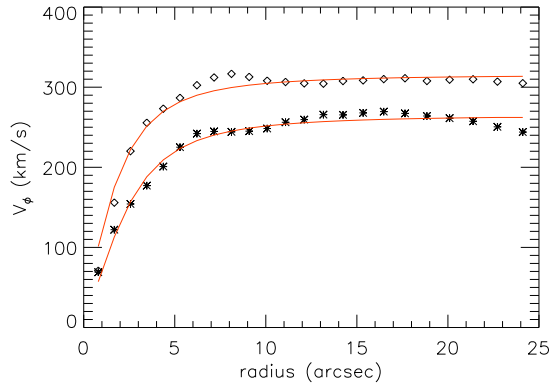,angle=0,width=8cm}}
\caption{Rotation curves of the ionised gas (open diamonds) and stars (stars), together with their best fit power-law models (red curves). The rotation curves have been extracted from the \sauron\ velocity fields.}
\label{fig:fit_vphi}
\end{figure}

To be able to fit the observed velocity we need to convolve our model
with the point-spread function (PSF) of the observations, and take the
binning into account that results from the finite pixelsize of the
CCD. We therefore constructed a two-dimensional velocity field of the
extracted rotation curve, such that

\begin{equation}
\label{eq:velfield}
V(R, \phi) = v_{\mathrm{mod}} \cos \phi \sin i,
\end{equation}

\noindent
and we convolved this field with a kernel as described in the appendix
of Qian et al. (1995)\nocite{1995MNRAS.274..602Q}. This kernel takes
into account the blurring caused by the atmosphere and the instrument
(FWHM = 1.4$^{\prime \prime}$, for the \sauron\ observations of
NGC~2974, see Emsellem et al.  2004\nocite{2004MNRAS.352..721E}) and
the spatial resolution of the reduced observations (0.8$^{\prime
  \prime}$ for \sauron). We extracted the velocity along the major
axis of the convolved velocity model and used the resulting rotation
curve to fit our observations.  The best fit is shown in
Figure~\ref{fig:fit_vphi}, and has a core radius $R_c = 2.1^{\prime
  \prime}$ ($\sim 0.2$ kpc).

Under the assumptions of Equation~(\ref{eq:vmod}), the asymmetric
drift correction of Equation~(\ref{eq:adc3}) reduces to

\begin{eqnarray}
\label{eq:adc_text}
V_c^2 = \overline{v_\phi}^2 - \sigma_R^2 \Big[ \frac{\partial \ln \Sigma}
{\partial \ln R} + \frac{\partial \ln \sigma_R^2}{\partial \ln R} + 
\frac{R^2}{2R_{\mathrm{mod}}^2} + \nonumber\\ \frac{\kappa R^2}
{\kappa(2R_{\mathrm{mod}}^2\!-\!R^2) + R^2} \Big],
\end{eqnarray}

\noindent
where $\overline{v_\phi}$ is the observed velocity, $\Sigma$ is the
surface brightness of the ionised gas and $\sigma_R$ the radial
dispersion of the gas. The last two terms in the equation are
connected to the shape of the velocity ellipsoid, with $\kappa$
indicating the alignment of the ellipsoid, see Appendix~\ref{sec:adc}.

\begin{figure}
    \centerline{\psfig{figure=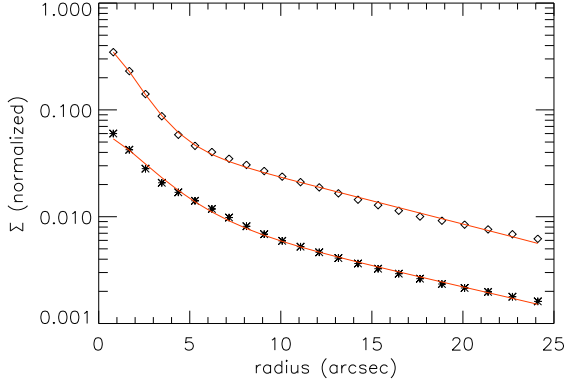,angle=0,width=8cm}}
\caption{Profile and fit to the surface brightness of the ionised gas (diamonds) and to the stars (stars). Both profiles have been normalized, and the profile of the stellar surface brightness has been offset by a factor of 10, to distinguish it from the gaseous one.}
\label{fig:dens_fit}
\end{figure}

To determine the slope of the surface brightness profile, we run
kinemetry on the \Oiii\ flux map, extracting the surface brightness
along ellipses with the same position angle and flattening as the
ones used to describe the velocity field. To decrease the noise we
fit a double exponential function to the profile,

\begin{equation}
\label{eq:disp}
\Sigma(R) = \Sigma_0 e^{-R/R_0} + \Sigma_1 e^{-R/R_1},
\end{equation}

\noindent
and determine the slope needed for the asymmetric drift correction
from this parametrisation. The observed surface brightness profile and
its fit are shown in Figure~\ref{fig:dens_fit}. As with the velocity
profile, we convolved our model of the surface brightness during the
fit with the kernel of Qian et al. (1995)\nocite{1995MNRAS.274..602Q}
to take seeing and sampling into account.

$\sigma_R$ can be obtained from the observed velocity dispersion
$\sigma$ using Equation~(\ref{eq:obs}). Along the major axis, and
under the assumptions made above, this expression simplifies to

\begin{equation}
\label{eq:sigma_simple}
\sigma_{\mathrm{obs}}^2 = \sigma_R^2 \Big[ 1 - \frac{R^2 \sin^2 i}{2R_{\mathrm{mod}}^2} -  \frac{R^2 \cos^2 i} {\kappa R_{\mathrm{mod}}^2(2-R^2/R_{\mathrm{mod}}^2) + R^2}  \Big],
\end{equation}

\noindent
with $R_{\mathrm{mod}}$ defined in Equation~(\ref{eq:rmod}), and
adopting $R_c = 2.1^{\prime \prime}$ from the velocity profile .

We choose $\kappa = 0.5$, which is a typical value for a disc galaxy
(e.g. Kent \& de Zeeuw 1991\nocite{1991AJ....102.1994K}), but we also
experimented with other values for this parameter. Varying $\kappa$
between 0 and 1 resulted in differences in $V_c$ of approximately 10
\kms, and we adopt this value into the error bars of our final
rotation curve.

To obtain the slope of $\sigma_R$ we follow the same procedure as for
the surface brightness, extracting the profile of
$\sigma_{\mathrm{obs}}$ from the velocity dispersion map with
kinemetry. We assume for the moment that turbulence is negligible in
the galaxy ($\sigma_{\mathrm{turb}} = 0$) and subtract quadratically
$\sigma_{\mathrm{thermal}} = 10$ \kms\ from $\sigma_{\mathrm{obs}}$.
We convert the resulting $\sigma_{\mathrm{obs}} =
\sigma_{\mathrm{grav}}$ into $\sigma_R$ using the relation in
Equation~(\ref{eq:sigma_simple}). We parametrise this profile by

\begin{equation}
\sigma_R(R) = \sigma_0 + \sigma_1 e^{-R_{\mathrm{mod}}/R_1}.
\end{equation}

\begin{figure}
    \centerline{\psfig{figure=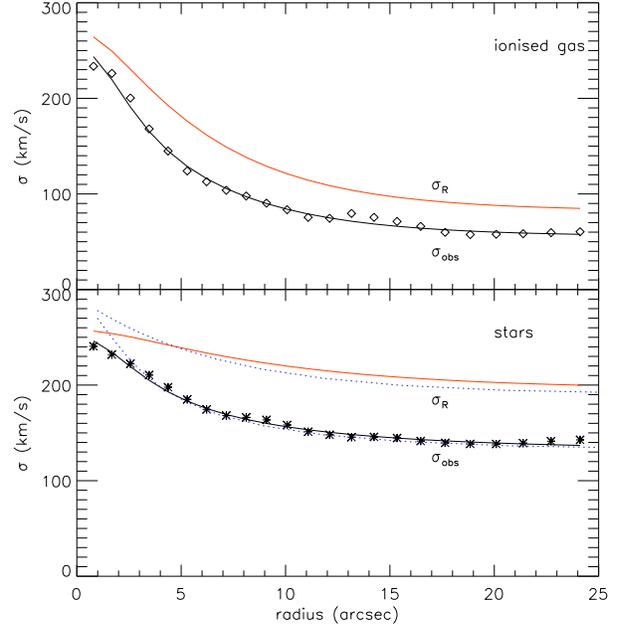,angle=0,width=8cm}}
\caption{Top panel: observed velocity dispersion $\sigma_{\mathrm{obs}}$ of the ionised gas (black diamond) with its best fit (black line). The red line denotes the radial dispersion $\sigma_R$, calculated from the fit to $\sigma_{\mathrm{obs}}$. Bottom panel: same as above, but now for the stellar observed velocity dispersion (black stars). The dotted blue lines represent $\sigma_R$ and $\sigma_{\mathrm{obs}}$ as extracted from the Schwarzschild model of Krajnovi\'c et al. (2005).}
\label{fig:sigma_fit}
\end{figure}

\noindent
This profile has a core in the centre (introduced by
$R_{\mathrm{mod}}$), so that we can better reproduce the flattening of
the profile towards the centre. Again, we convolved our model to take
seeing and sampling into account during the fit. The top panel of
Figure~\ref{fig:sigma_fit} shows the resulting profile and fit, as
well as the observed velocity dispersion.

\begin{figure}
    \centerline{\psfig{figure=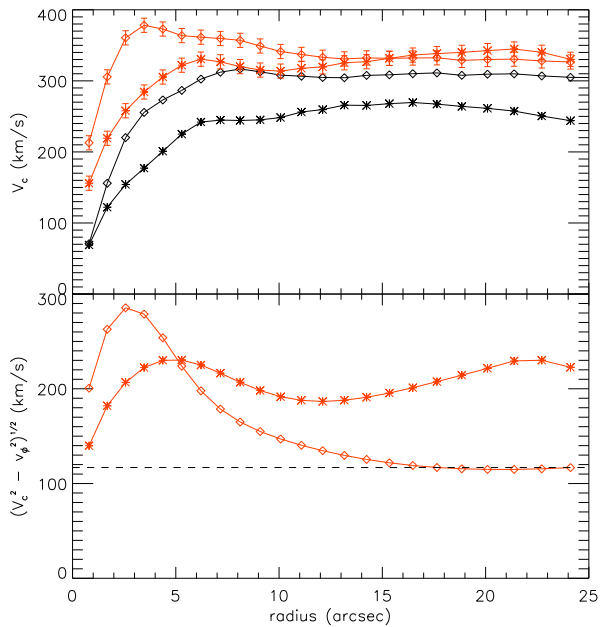,angle=0,width=8cm}}
\caption{Top panel: observed rotation curve of the ionised gas (black diamonds) and stars (black stars) with their asymmetric drift corrected curve (red diamonds and stars). The correction to the ionised gas seems too high in the central part of the galaxy, when compared to the corrected stellar rotation curve. Bottom panel: asymmetric drift correcion $(V_c^2 - v_\phi^2)^{1/2}$ of the ionised gas (diamonds) and the stars (stars). The dashed line denotes the mean asymmetric drift correction of the ionised gas outside 15 arcseconds.}
\label{fig:asym}
\end{figure}

We first assume that turbulence plays no role in this galaxy, and we use
$\sigma_R$ as computed above to calculate the asymmetric drift
correction (Equation~\ref{eq:adc_text}). The resulting rotation curve,
as well as the observed rotation curve of the ionised gas, is shown in
the top panel of Figure~\ref{fig:asym}.

To check our asymmetric drift corrected rotation curve of the ionised
gas, we compare it with the asymmetric drift corrected stellar
rotation curve. Stars do not feel turbulence and are not influenced by
thermal motions like the gas, and therefore their observed velocity
dispersion contains only contributions of gravitational interactions:
$\sigma_{\mathrm{obs}} = \sigma_{\mathrm{grav}}$. If we are correct
with our assumption that turbulence does not play a role in the
ionised gas, then the stellar corrected rotation curve should overlap
with the corrected curve of the gas. If it does not, then we know that
we should not have neglected the turbulence.

To derive the asymmetric drift correction of the stars, we obtain
the observed rotation curve, surface density and velocity dispersion
of the stars from our \sauron\ observations with kinemetry, and
parametrise them in the same way as we did for the ionised gas (see
Figures ~\ref{fig:fit_vphi} - \ref{fig:sigma_fit} for the observed
profiles and their models). The models were convolved during the
fitting as described for the ionised gas. Because for the stars
$\sigma_{\mathrm{obs}} = \sigma_{\mathrm{grav}}$ we do not need to
subtract $\sigma_{\mathrm{thermal}}$ as we did for the ionised gas and
hence can calculate $\sigma_R$ directly from
Equation~(\ref{eq:sigma_simple}), where we inserted a core radius $R_c
= 3.0^{\prime \prime}$ from the stellar velocity model.

In the above, we assumed that the stars lie in a thin disc, which is
not the case in NGC~2974. To check the validity of our thin disc
approximation for our model of $\sigma_R$, we extract this quantity
from the Schwarzschild model of Krajnovi\'c et al.
(2005)\nocite{2005MNRAS.357.1113K}, for $\theta=84^\circ$, close to
the $z=0$ plane. The resulting profile is smoothed and shown as the
upper dotted blue line in Figure~\ref{fig:sigma_fit}. It is not a fit
to the data, but derived independently from the Schwarzschild model,
and agrees very well with the stellar $\sigma_R$ we got from
kinemetry. Also, $\sigma_{\mathrm{obs}}$ derived from the
Schwarzschild model (lower dotted blue line) agrees with the results
from kinemetry, giving us confidence that our stellar $\sigma_R$ is
reliable.

When we compare the asymmetric drift corrected rotation curves of the
ionised gas and of the stars in Figure~\ref{fig:asym}, then it is
clear that although for $R > 15^{\prime \prime}$ the agreement between
the curves is very good, the correction for the gas is too high in the
central part of the galaxy.  This is an indication that turbulence
cannot be neglected here, and needs to be taken into account.

\subsection{Turbulence}
\label{sec:turb}

\begin{figure}
    \centerline{\psfig{figure=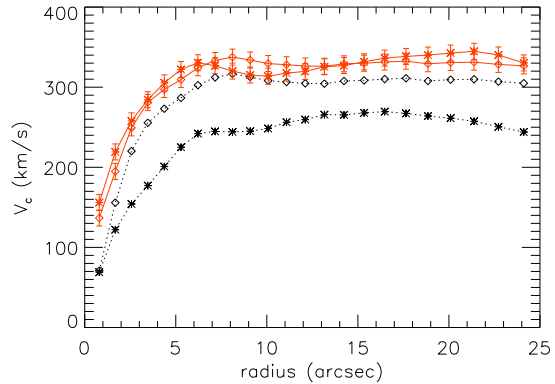,angle=0,width=8cm}}
\caption{Asymmetric drift corrected rotation curve of the gas (red diamonds), removing turbulence as described in the text. The red stars denote the asymmetric drift corrected rotation curve of the stars. The two curves agree very well, suggesting that our turbulence model is adequate for our purposes. For comparison, also the observed rotation curves of the gas and stars are plotted (black diamonds and stars, respectively).}
\label{fig:asym_stars}
\end{figure}

For radii larger than $15^{\prime \prime}$, the corrected velocity
curve of the ionised gas is in agreement with the stellar corrected
velocity curve, and since stellar motions are not influenced by
turbulence, we can conclude that in this region turbulence is
negligible. The bottom panel of Figure~\ref{fig:asym} shows the
asymmetric drift correction $(V_c^2 - v_\phi^2)^{1/2}$ itself, and we
see that outside 15$^{\prime \prime}$, the correction is more or less
constant at approximately 120 \kms (dashed line). In order to remove the turbulence
from the central region in NGC~2974, we now assume that the asymmetric
drift correction has the same value everywhere in the galaxy, namely
120 \kms. We add this value quadratically to the observed rotation
curve of the ionised gas, and obtain the rotation curve shown in
Figure~\ref{fig:asym_stars}. This corrected rotation curve agrees
strikingly well with the corrected rotation curve of the stars, and
this is a strong indication that our model for turbulence is
reasonable, and at least good enough to get a reliable rotation curve
for the ionised gas.

We now investigate the random motions resulting from turbulence and
gravitational interaction in some more detail. Since we assumed a
constant asymmetric drift correction $(V_c^2 - v_\phi^2)^{1/2}$ of
$\sim 120$ \kms, we can at each radius calculate the corresponding
$\sigma_R$ with Equation~(\ref{eq:adc_text}). Using
Equation~(\ref{eq:sigma_simple}) we obtain the observed velocity
dispersion, which in this case consists only of
$\sigma_{\mathrm{grav}}$. Since we know $\sigma_{\mathrm{obs}}$, we
can subtract quadratically $\sigma_{\mathrm{grav}}$ and
$\sigma_{\mathrm{thermal}}$ = 10 \kms\ to obtain
$\sigma_{\mathrm{turb}}$.

\begin{figure}
    \centerline{\psfig{figure=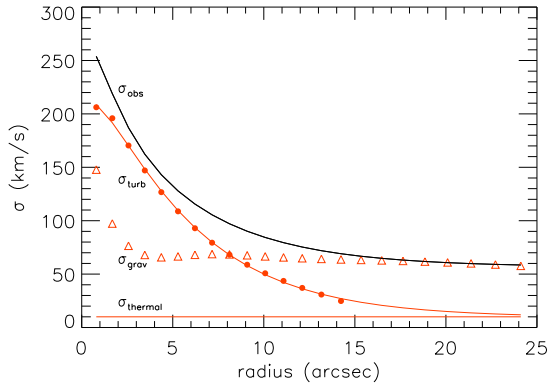,angle=0,width=8cm}}
\caption{Observed deconvolved velocity dispersion (solid black line), with its components $\sigma_{\mathrm{thermal}}$ (red horizontal line), $\sigma_{\mathrm{grav}}$ (open red triangles) and $\sigma_{\mathrm{turb}}$ (filled red dots). An exponential fit to $\sigma_{\mathrm{turb}}$ is overplotted.}
\label{fig:sigma_turb}
\end{figure}

Figure~\ref{fig:sigma_turb} shows $\sigma_{\mathrm{obs}}$ (deconvolved
model) and its components $\sigma_{\mathrm{thermal}}$,
$\sigma_{\mathrm{grav}}$ and $\sigma_{\mathrm{turb}}$.  We fitted a
single exponential function (Equation~\ref{eq:disp}) with $R_c =
2.1^{\prime \prime}$ to the inner 15 arcseconds of
$\sigma_{\mathrm{turb}}$ and find that with this parametrisation we
can get a decent fit. We find a lengthscale of 5.0$^{\prime \prime}$
for the turbulence. The fit is also shown in
Figure~\ref{fig:sigma_turb}. 
  
\section{Mass model and dark matter content}

In this section we combine the corrected rotation curve of the ionised
gas with the rotation curve of the neutral gas. The rotation curve of
NGC~2974 rises quickly to a maximal velocity and then declines to a
somewhat lower velocity, after which it flattens out (see e.g.
Figure~\ref{fig:maxdisc}). Unfortunately, we lack the data to study
this decline in more detail, because our \HI\ ring is not filled. The
behaviour of our rotation curve is similar to what is seen in other
bright galaxies with a concentrated light distribution (Casertano \&
van Gorkom 1991\nocite{1991AJ....101.1231C}, Noordermeer et al.
2007\nocite{2007MNRAS.376.1513N}). The decline of the rotation curve
in such systems could indicate that the mass distribution in the
centre is dominated by the visible mass and that the dark halo only
takes over at larger radii. In contrast, in galaxies where the light
distribution is less concentrated, such as low-luminosity later-type
galaxies, the rotation curves does not decline (e.g. Spekkens \&
Giovanelli 2006\nocite{2006AJ....132.1426S}, Catinella, Giovanelli \&
Haynes 2006\nocite{2006ApJ...640..751C}).

We separately model the contribution of the stars, neutral gas and
dark halo to the gravitational potential. Also we derive the total
mass-to-light ratio as a function of radius, and obtain a lower limit
on the dark matter fraction in NGC~2974.

In our model, we do not take the weak bar system of NGC~2974 into
account. Emsellem et al. (2003)\nocite{2003MNRAS.345.1297E} find that
the perturbation of the gravitational potential caused by the inner
bar in their model of this galaxy is less than 2 per cent.  Also, we
find that the harmonic coefficients that could be influenced by a
large scale bar ($s_1$, $s_3$ and $c_3$) are small compared to the
dominant term $c_1$ ($<$ 4 per cent). We therefore conclude that
although the rotation curve probably is affected by the presence of
the bar system, this effect is small, and negligible compared to the
systematic uncertainties introduced by the asymmetric drift
correction. Furthermore, the largest constraints in our models come
from the rotation curve at large radii, where we showed that the
elongation of the potential is consistent with axisymmetry.

\subsection{Stellar contribution}

\begin{table}
\begin{center}
\begin{tabular}{l|c|c}
\hline\hline
  & HST/WFPC2 & MDM \\
\hline
Filter Band &   F814W & $I$ \\  
Exposure Time (s) &    250 & 1500 \\ 
Field of View  &   $32^{\prime \prime} \times 32^{\prime \prime}$ & $17.4^{\prime} \times 17.4^{\prime}$    \\  
Pixel scale (arcsec) &     0.0455  & 0.508 \\ 
Date of Observation   &  16 April 1997  &  26 March 2003 \\  
\hline
\end{tabular}
\end{center}
\caption{Properties of the space- and ground-based imaging of NGC~2974, used to model the stellar contribution to the potential. The MDM image was constructed of 3 separate exposures, resulting in a total integration time of 1500 s.} 
\label{tab:images}
\end{table}

The contribution of the stellar mass to the gravitational potential
and the corresponding circular velocity can be obtained by
deprojecting and modelling the surface photometry of the galaxy. We
use the Multi-Gaussian Expansion (MGE) method for this purpose, as
described in Cappellari (2002)\nocite{2002MNRAS.333..400C}.

Krajnovi\'c et al. (2005)\nocite{2005MNRAS.357.1113K} presented an MGE
model of NGC~2974, based upon the PC part of a dust-corrected
WFPC2/F814W image and a ground-based $I$-band image obtained at the
1.0m Jacobus Kapteyn Telescope (JKT). This image was however not deep
enough to yield an MGE model that is reliable out to 5 $R_e$ or
120$^{\prime \prime}$, which is the extent of our rotation curve. We
therefore construct another MGE model, replacing the JKT $I$-band
image with a deeper one obtained with the 1.3-m McGraw-Hill Telescope
at the MDM Observatory (see Table~\ref{tab:images}). This image is
badly contaminated by a bright foreground star, so we do not include
the upper half of the image in the fit. Since our model is
axisymmetric, enough signal remained to get a reliable fit.  We also
exclude other foreground stars and bleeding from the image. The
parameters of the point spread function (PSF) for the WFPC2 image were
taken from Krajnovi\'c et al. (2005)\nocite{2005MNRAS.357.1113K}.

\begin{figure}
\centerline{\psfig{figure=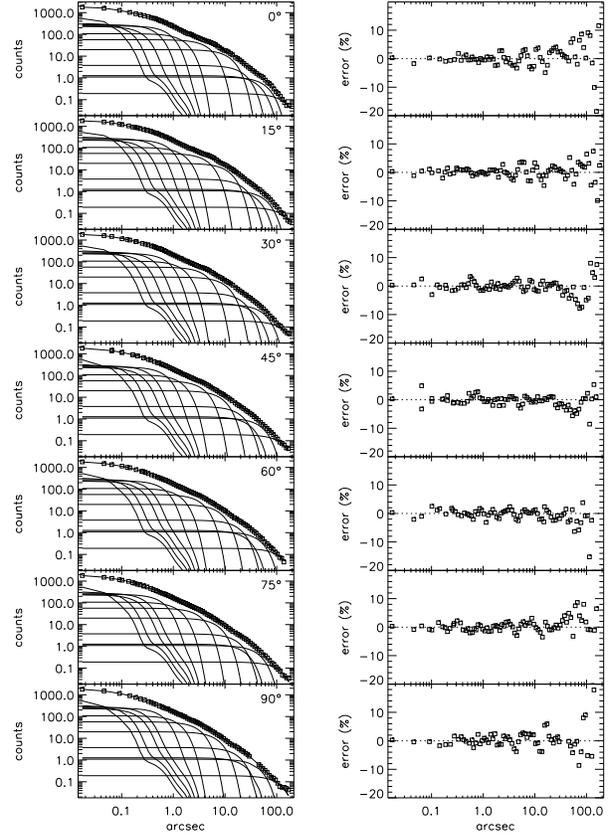,angle=0,width=8cm}}
\caption{Left panels: comparison between the WFPC2 and MDM photometry (open squares) and the convolved gaussians composing the MGE model of NGC~2974 (solid line), as a function of radius. Different panels show different angular sectors.  Right panels: relative error of the MGE model compared to the data, as a function of radius.}
\label{fig:mge_res}
\end{figure}

We match the ground-based MDM image to the higher resolution WFPC2
image, and use it to constrain the MGE-fit outside 15$^{\prime
  \prime}$. Outside 200$^{\prime \prime}$, the signal of the galaxy
dissolves into the background and we stop the fit there. We are
therefore confident of our MGE model out to a radius of at least
120$^{\prime \prime}$, which is the extent of the observed \HI\ 
rotation curve.  The goodness of fit can be examined as a function of
radius in Figure~\ref{fig:mge_res}.

We forced the axial ratios $q_j$ of the Gaussians to lie in the
interval [0.58, 0.80] (which is the same range as Krajnovi\'c et al.
(2005)\nocite{2005MNRAS.357.1113K} used in their paper), maximising the
number of allowed inclinations and staying as close as possible to a
model with constant ellipticity, without significantly increasing the
$\chi^2$ of the fit. This resulted in an MGE model consisting of
twelve Gaussians, whose parameters can be found in
Table~\ref{tab:mge}.  The parameters of the inner Gaussians agree very
well with the ones in Krajnovi\'c et al.'s model, which
is not surprising as we used the same dust-corrected WFPC image. The
outer Gaussians deviate, where their JKT image is replaced by our MDM
image.

\begin{figure}
\begin{tabular}{c}
\centerline{\psfig{figure=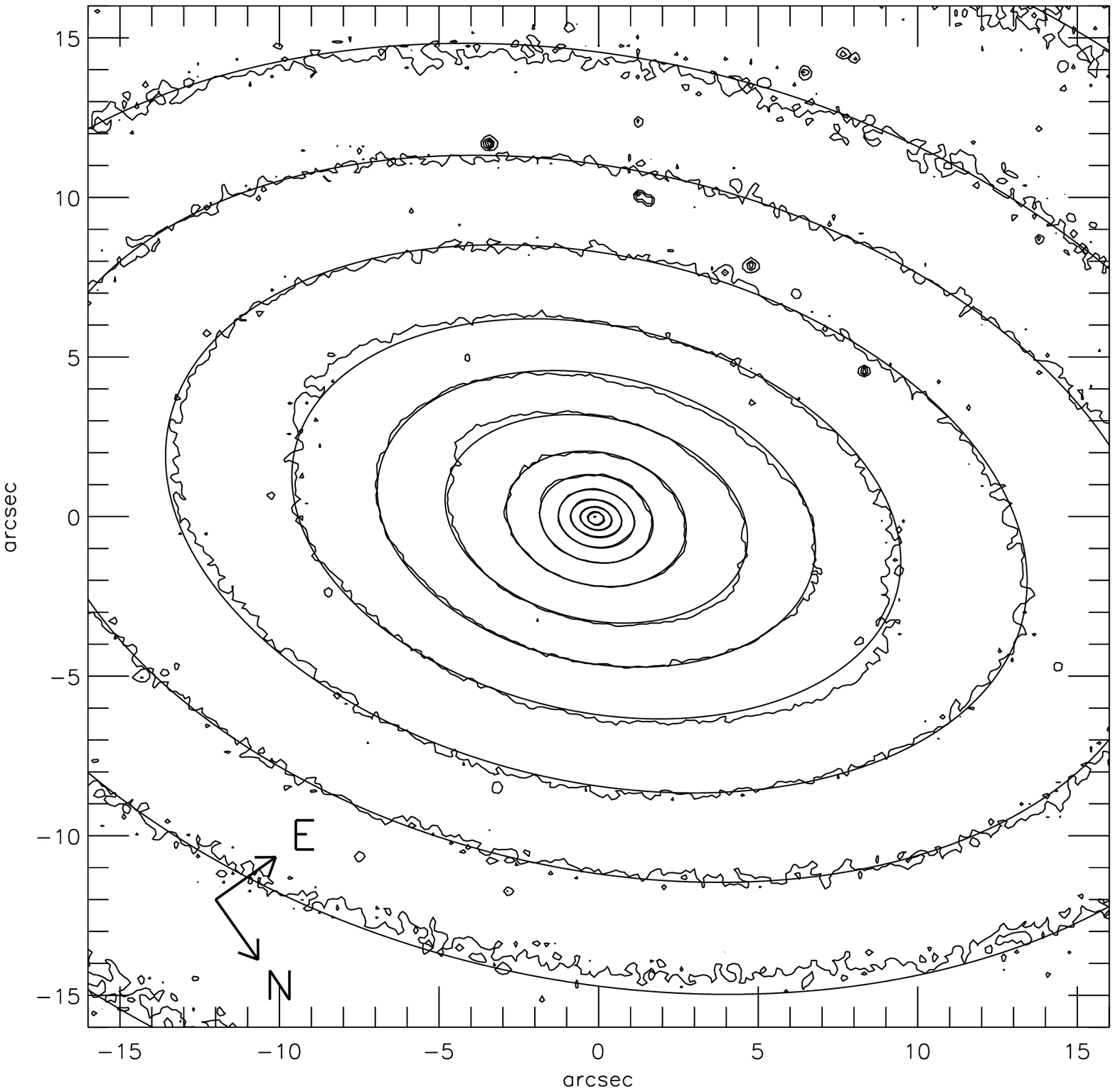,angle=0,width=7.8cm}} \\
\centerline{\psfig{figure=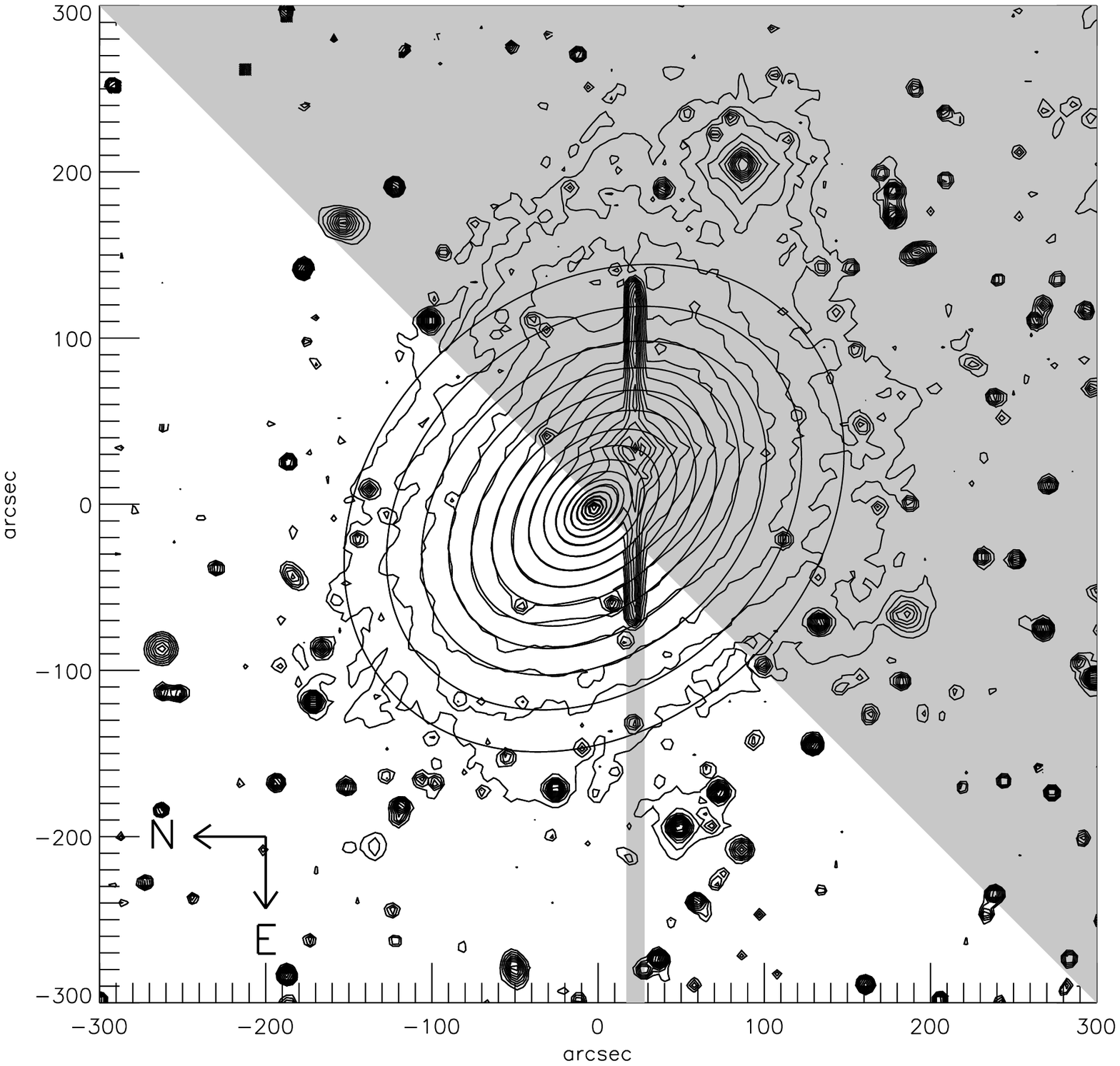,angle=0,width=7.8cm}} 
\end{tabular}
\caption{Contour maps of the $I$-band photometry of NGC~2974. From top to bottom: dust-corrected PC of WFPC2/F814W image and MDM image. The grey area in the MDM image indicates the area that has been excluded from the fit, because of contamination by the bright foreground star. Apart from this area, other foreground star were also masked during the fit. Overplotted are the contours of the MGE surface brightness model, convolved with the PSF of WFPC2.} 
\label{fig:mge_contours}
\end{figure}

Figure~\ref{fig:mge_contours} shows the WFPC2 and MDM photometry and
the overlaid contours of the MGE model. Also shown is the masked MDM
image. The deviations in the WFPC plot between the isophotes and the
MGE model around 10$^{\prime \prime}$ are point-symmetric and
therefore probably reminiscent of a spiral structure (e.g. Emsellem et
al.  2003\nocite{2003MNRAS.345.1297E}). The deviations are however
small, and we conclude that the MGE model is a good representation of the
galaxy surface brightness.

\begin{table}
\begin{center}
\begin{tabular}{l|c|c|c|c}
\hline\hline
$j$ & $I_j(L_\odot$ pc$^-2)$ & $\sigma_j$ (arcsec) & $q_j$ & $L_j (
\times 10^9 L_\odot )$ \\
\hline
1 &     187628. &   0.0376306 &    0.580000 & 0.0099 \\
2 &     44798.9 &   0.0923231 &    0.800000 &   0.0197 \\
3 &     25362.4 &    0.184352 &    0.800000 &   0.0445 \\
4 &     28102.0 &    0.343100 &    0.586357 &    0.1251 \\
5 &     23066.0 &    0.607222 &    0.722855 &    0.3964 \\
6 &     9694.88 &     1.20984 &    0.774836 &    0.7089 \\
7 &     5019.87 &     3.56754 &    0.659952 &     2.7186 \\
8 &     1743.48 &     9.23267 &    0.580000 &     5.5578 \\
9 &     329.832 &     16.9511 &    0.770081 &     4.7057 \\
10 &    111.091 &      30.5721 &    0.580000 &     3.8829 \\
11 &    96.2559 &      44.0573 &    0.717554 &     8.6440 \\
12 &    16.7257 &      103.085 &    0.800000 &     9.1678 \\
\hline
\end{tabular}
\end{center}
\caption{Parameters of the Gaussians of the MGE model of NGC~2974. From left to right: number of the Gaussian, central intensity, width (standard deviation), axial ratio and total intensity.}
\label{tab:mge}
\end{table}

\subsection{Gas contribution}

The contribution of the \HI\ ring to the gravitational potential is
small compared to the stars and halo ($5.5 \times 10^8 M_\odot$, three
orders of magntiude smaller than the stellar mass) but still included
in our mass models. We include a factor 1.3 in mass to account for the
helium content of the ring. The mass of the ionised gas is estimated
at only $2.2 \times 10^5 M_\odot$ (Sarzi et al.
2006\nocite{2006MNRAS.366.1151S}), and therefore can be neglected in
our models.

\subsection{Mass-to-light ratio}

By comparing the observed rotation curve and the light distribution
from the MGE model, we can already calculate the mass-to-light ratio
in NGC~2974. The enclosed mass within a certain radius $r$ in a
spherical system follows directly from the circular velocity:

\begin{equation}
\label{eq:mass_vc}
M(<r) = \frac{V_c^2 r}{G},
\end{equation}

\noindent
with $G$ the gravitational constant. Here we assume that the
gravitational potential of the total galaxy is spherical symmetric.
This is clearly not the case for the neutral gas, which resides in a
thin disc. However, the total mass of the gas is three orders of
magnitudes smaller than the total mass, and therefore can be
neglected. Also, the stars reside in a flattened potential, as can be
shown from their MGE model. But since we cannot disentangle the
contributions of the stars and the dark matter to the observed
rotation velocity {\it a priori}, we will for the moment assume that
also the stellar mass density can be approximated by a spherical
distribution.

Since we know the mass within a sphere of radius $r$, we also need to
calculate the enclosed $I$-band luminosity within a sphere. We first
obtain the gravitational potential of our MGE model as a function of
radius (see appendix A of Cappellari et al.
2002\nocite{2002ApJ...578..787C}). Here, we take the flattening of the
separate Gaussians into account. We subsequently calculate the
corresponding circular velocity, with an arbitrary $M_*/L$. To find
the luminosity enclosed in a sphere we calculate the spherical mass
needed to produce this circular velocity with
Equation~(\ref{eq:mass_vc}), and convert this mass back to a
luminosity using the same $M_*/L$ that we used to calculate the
velocity curve. This way we have replaced the luminosity within a
flattened axisymmetric ellipsoid (oblate sphere) by a sphere with
radius equal to the long axis of the ellipsoid.

With this method we arrive at a mass-to-light ratio $M/L_I=$ 8.5
$M_\odot/L_{\odot,I}$ at 5 effective radii ($1R_e = 24^{\prime
  \prime}$).  In the literature, this value is usually expressed in
$B$-band luminosities.  Using an absolute magnitude of $M_B= -20.07$
for NGC~2974 (see Table~\ref{tab:ngc2974}), we find that $M/L_B=$ 14
$M_\odot/L_{\odot,B}$.  We checked that $M_B$ is consistent with our
MGE model, adopting a colour $B-I = 2.13$ for NGC~2974 (see Tonry et
al. 2001\nocite{2001ApJ...546..681T} and Table~\ref{tab:ngc2974}).
\HI\ studies of other early-type galaxies yield similar numbers
(Morganti et al.  1997\nocite{1997AJ....113..937M} and references
therein). For example, Franx et al.
(1994)\nocite{1994ApJ...436..642F} find $M/L_B = $16 $
M_\odot/L_{\odot,B}$ at 6.5 $R_e$ using the \HI\ ring around IC~2006,
and Oosterloo et al. (2002)\nocite{2002AJ....123..729O} report $M/L_B
=$ 18 $M_\odot/L_{\odot,B}$ for NGC~3108 at 6 $R_e$.

\begin{figure}
    \centerline{\psfig{figure=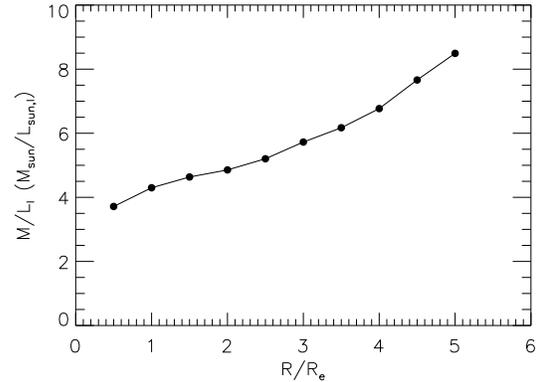,angle=0,width=8cm}}
\caption{$M/L_I$ as a function of radius. The increase of $M/L_I$ is a strong indication for a dark matter halo around NGC~2974.}
\label{fig:ml}
\end{figure}

Figure~\ref{fig:ml} shows the increase of $M/L_I$ with radius. We
find that within 1 $R_e$, $M/L_I =$ 4.3 $ M_\odot/L_{\odot,I}$, which
agrees with the results from Schwarzschild modeling of Krajnovi\'c et
al. (2005)\nocite{2005MNRAS.357.1113K} and Cappellari et al.
(2006)\nocite{2006MNRAS.366.1126C}. The increase of $M/L$ indicates
that the fraction of dark matter grows towards larger radii.

\subsection{Dark matter fraction}

\begin{figure}

  \centerline{\psfig{figure=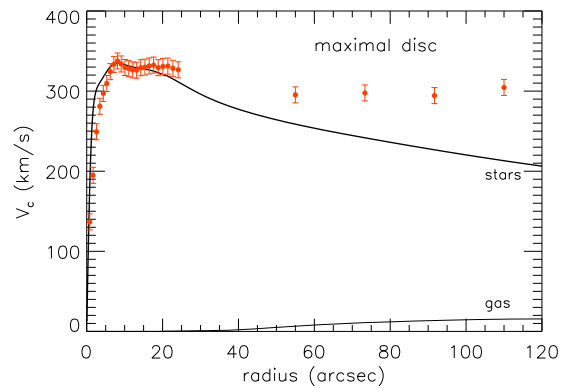,angle=0,width=8cm}}
\caption{Best fit of a maximal disc model to the observed rotation curve. Red points indicate the observations and the thick black curve is the fit to these datapoints. This curve is calculated from the combined stellar and gaseous mass, and convolved with a kernel that takes seeing and sampling into account. The stellar mass-to-light ratio in this model is 3.8 $M_\odot/L_{\odot,I}$.}
\label{fig:maxdisc}
\end{figure}

To calculate the dark matter fraction, we need to know the stellar
mass-to-light ratio $M_*/L$.  An upper limit on $M_*/L_I$ can be
derived by constructing a maximal disc model. From the MGE model we
calculate a rotation curve (taking the flattening of the potential
into account, as in Cappellari et al.
2002\nocite{2002ApJ...578..787C}), and we increase $M_*/L_I$ until the
calculated curve exceeds the observed rotation curve. This way, we
find that $M_*/L_I$ cannot be larger than 3.8 $M_\odot/L_{\odot,I}$.
We plotted the rotation curve of the maximal disc model, together with
the observed rotation curve in Figure~\ref{fig:maxdisc}. The rotation
curve of the model has been convolved to take seeing and the resolution
of the observations into account, as described in
\S~\ref{sec:adcshort}. The contribution of the neutral gas to the
gravitational potential has been included in the model, but has only a
negligible effect on the fit. 

It is clear that even in the maximal disc model, a dark matter halo is
needed to explain the flat rotation curve of the \HI\ gas at large
radii. From this model, we can calculate a lower limit to the dark
matter fraction in NGC~2974. We then find that within one $R_e$, 12
per cent of the total mass is dark, while within 5 $R_e$, this
fraction has grown to 55 per cent.

There is however no reason to assume that the stellar mass-to-light
ratio is well represented by its maximal allowed value. Cappellari et
al.  (2006)\nocite{2006MNRAS.366.1126C} find $M_*/L_I = $ 2.34
$M_\odot/L_{\odot,I}$ for NGC~2974, measured from line-strength values
using single stellar population models. The formal error that they
report on this mass-to-light ratio is $\sim$ 10 per cent, but they
warn that this value is strongly assumption dependent. Secondary star
formation in a galaxy can result in an underestimation of $M_*/L$, and
the GALEX observations of Jeong et al.
(2007)\nocite{2007MNRAS.376.1021J} indeed show evidence for recent
star formation in NGC~2974. The population models of Cappellari et al.
(2006)\nocite{2006MNRAS.366.1126C} are based on a Kroupa initial mass
function (IMF), but if instead a Salpeter IMF is used, their $M_*/L_I$
values increase by $\sim$ 40 per cent, which for NGC~2974 would result
in $M_*/L_I = $ 3.3 $M_\odot/L_{\odot,I}$.  Cappellari et al.
(2006)\nocite{2006MNRAS.366.1126C} discard the Salpeter IMF based
models, because for a large part of their sample their models then
have $M_*/L_I > M_{\mathrm{tot}}/L_I$, which is unphysical.

\begin{figure}
    \centerline{\psfig{figure=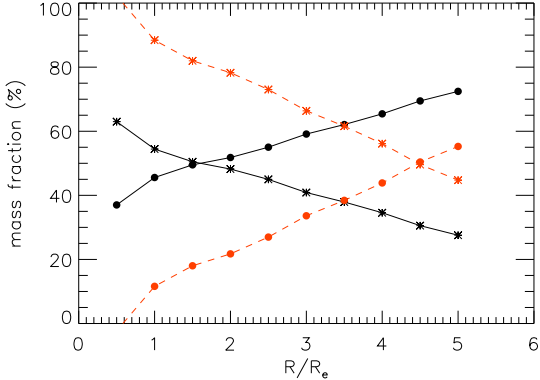,angle=0,width=8cm}}
\caption{Dark matter fraction (filled dots) and stellar mass fraction (stars). The black solid lines assume a stellar $M/L_I$ of 2.34 $M_\odot/L_{\odot,I}$ as predicted by single stellar population models of NGC~2974. The red dashed lines provide lower and upper limits for the dark matter and stellar mass fraction, respectively,  and are based on $M_*/L_I$ = 3.8 $M_\odot/L_{\odot,I}$, from the maximal disc model. }
\label{fig:massfrac}
\end{figure}

If we adopt $M_*/L_I =$  2.34 $M_\odot/L_{\odot,I}$ from the
stellar population models, then 46 per cent of the total mass within 1
$R_e$ is dark.  The dark matter fraction increases to 72 per cent
within 5 $R_e$. See Figure~\ref{fig:massfrac} for the change in dark
matter fraction as a function of radius, and the comparison with the
lower limits derived above.

Gerhard et al. (2001)\nocite{2001AJ....121.1936G} and Cappellari et
al. (2006)\nocite{2006MNRAS.366.1126C} find an average dark matter
fraction of $\sim$ 30 per cent within one effective radius in
early-type galaxies, but we note that NGC~2974 is an outlier in the
sample of Cappellari et al. The value of 47 per cent that we find is a
bit high compared to this average, though the minimal fraction of dark
matter is 14 per cent in our galaxy. Without an accurate determination
of $M_*/L$ we can not give a more precise estimate on the dark matter
fraction in NGC~2974.

\subsection{Halo models}

We now include a dark halo in our model, to explain the flat rotation
curve that we extracted from the \HI\ ring. We explore two different
halo models: the pseudo-isothermal sphere and the NFW profile.

The pseudo-isothermal sphere has a density profile given by:

\begin{equation}
\label{eq:rho_iso}
\rho(r) = \frac{\rho_0}{1+(r/r_c)^2},
\end{equation}

\noindent
where $\rho_0$ is the central density of the sphere, and $r_c$ is the
core radius. 

The velocity curve resulting from the density profile of the
pseudo-isothermal sphere is straightforward to derive analytically,
and given by

\begin{equation}
\label{eq:vc_iso}
V_c^2(r) = 4\pi G \rho_0 r_c^2 \Big( 1 - \frac{r_c}{r}\arctan{ \frac{r}{r_c}}\Big).
\end{equation}

The NFW profile was introduced by Navarro et
al. (1996)\nocite{1996ApJ...462..563N} to describe the haloes
resulting from simulations, taking a cold dark matter cosmology into
account. This profile has a central cusp, in contrast to the
pseudo-isothermal sphere which is core-dominated. Its density profile
is given by

\begin{equation}
\label{eq:rho_nfw}
\rho(r) = \frac{\rho_s}{r/r_s \big[1 + \big(\frac{r}{r_s}\big)^2\big]},
\end{equation}

\noindent
with $\rho_s$ the characteristic density of the halo and $r_s$ a
characteristic radius. The velocity curve of the NFW halo is given by

\begin{equation}
\label{eq:vc_nfw}
V_c^2(r) = V_{200}^2 \frac{\ln(1+cx) -cx/(1+cx)}{x[\ln(1+cx) - c/(1+c)]},
\end{equation}

\noindent
where $x = r/r_{200}$ and $c$ the concentration parameter defined by
$c=r_{200}/r_s$. $r_{200}$ is defined such that within this radius the
mean density is 200 times the cricital density $\rho_{\mathrm{crit}}$,
and $V_{200}$ is the circular velocity at that radius. These
parameters depend on the assumed cosmology.

We construct mass models of NGC~2974 including a dark matter halo with
the observed stellar and gaseous mass. We then calculate the circular
velocity resulting from our models, by adding the circular velocities
resulting from the separate components:

\begin{equation}
\label{eq:vctot}
V_c^2(r) = V_{c,\mathrm{halo}}^2+ V_{c,\mathrm{stars}}^2+ V_{c,\mathrm{gas}}^2, 
\end{equation}

\noindent
and fit these to our observed rotation curve. The inner 25$^{\prime
  \prime}$ of our model rotation curve, which are based on the
\sauron\ ionised gas measurements, are convolved with a kernel to take
seeing and sampling into account, as described in
\S~\ref{sec:adcshort}. 

For both profiles, we found that we could not constrain the stellar
mass-to-light ratio in our models because of degeneracies: for each
$M_*/L_I$ below the maximal disc value of 3.8 $M_\odot/L_{\odot,I}$ we
could get a decent fit. We therefore show two fits for each model,
with $M_*/L$ values that are justified by either linestrength
measurements and single stellar population models ($M_*/L_I = $ 2.34
$M_\odot/L_{\odot,I}$) or the observed rotation curve itself ($M_*/L_I
= $ 3.8 $M_\odot/L_{\odot,I}$). This last case would be a model
requiring a minimal halo.

\begin{figure}
    \centerline{\psfig{figure=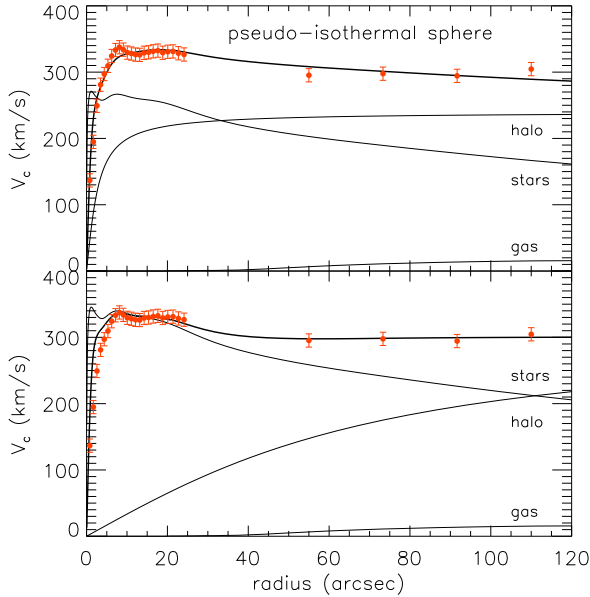,angle=0,width=8cm}}
\caption{Best fit models of a dark halo represented by a pseudo-isothermal sphere. The top panel has a stellar $M_*/L_I$ of 2.34 $M_\odot/L_{\odot,I}$ from stellar population models, and the bottom panel has $M_*/L_I =$ 3.8 $M_\odot/L_{\odot,I}$ from the maximal disc model.  The red dots are our observations from the ionised gas (asymmetric drift corrected) and \HI\ gas. The rotation curves resulting from the potentials  of the halo, stars and gas are plotted separately, where the first two are unconvolved. The bold line denotes the fit to the data, and is the convolved rotation curve resulting from the combined potential of halo, stars and gas.}
\label{fig:iso}
\end{figure}

The best fit models for a dark halo described by a pseudo-isothermal
sphere is shown in Figure~\ref{fig:iso}. The model in the top panel
has a fixed $M_*/L_I =$ 2.34 $M_\odot/L_{\odot,I}$, while the bottom
panel shows the model with $M_*/L_I =$ 3.8 $M_\odot/L_{\odot,I}$.  The
first model fits the \sauron\ measurement of the rotation curve well,
but has a small slope at the outer part, where the observations show a
flat rotation curve.  Nevertheless, this model provides a good fit,
with a minimal $\chi^2 = 27$ for $27-2 = 25$ degrees of freedom. We
find for this model $\rho_0 =$ 19 $M_\odot$ pc$^{-3}$ and core radius
$r_c = 2.3^{\prime \prime} =$ 0.23 kpc. The second model with $M_*/L_I =
$ 3.8 $M_\odot/L_{\odot,I}$ provides a better fit to the \HI\ 
measurements, but has problems fitting the central part of the
rotation curve. The model has a lower central density $\rho_0 =$ 0.06
$M_\odot$ pc$^{-3}$ and larger core radius $r_c = 54^{\prime \prime} =$
5.4 kpc. The fit is worse than for the previous model, with $\chi^2 =
133$.

\begin{figure}
    \centerline{\psfig{figure=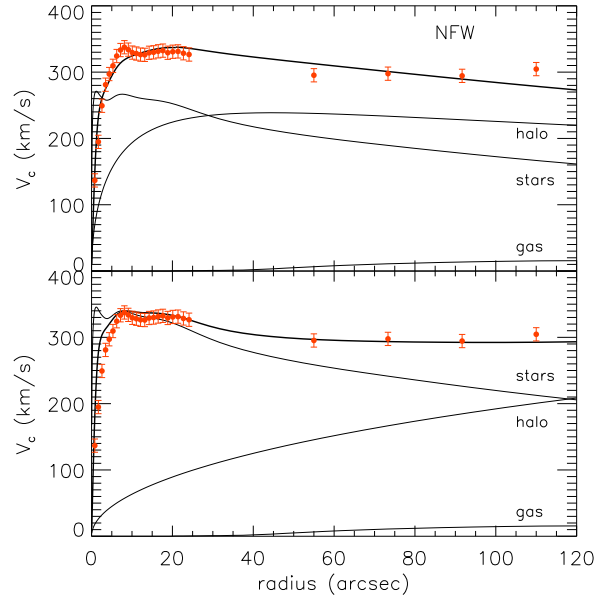,angle=0,width=8cm}}
\caption{Same as Figure~\ref{fig:iso}, but now with a dark halo contribution given by an NFW profile. The top panel has the stellar $M_*/L_I$ value from population models (2.34 $M_\odot/L_{\odot,I}$), and the bottom panel from the maximal disc model (3.8 $M_\odot/L_{\odot,I}$).}
\label{fig:nfw}
\end{figure}

Figure~\ref{fig:nfw} shows the best fitting-models with an NFW dark
halo. This model fits the data less well than the pseudo-isothermal
sphere: for the model with $M_*/L_I =$ 2.34 $M_\odot/L_{\odot,I}$ (top
panel) we find a minimal $\chi^2 = 44$ for $27-2$ degrees of freedom.
The corresponding parameters of the density function are $\rho_s =$
1.1 $M_\odot$ pc$^{-3}$ and $r_s = 21^{\prime \prime}$ = 2.1 kpc. For
$M_*/L_I =$ 3.8 $M_\odot/L_{\odot,I}$ the fit is worse ($\chi^2 =
144$) but the outer part of the rotation curve is better fitted. We
find $\rho_s =$ 1.1 $\times 10^{-3}$ $M_\odot$ pc$^{-3}$ and $r_s
\approx 1300^{\prime \prime}$, which corresponds to approximately 130
kpc.

\begin{figure}
    \centerline{\psfig{figure=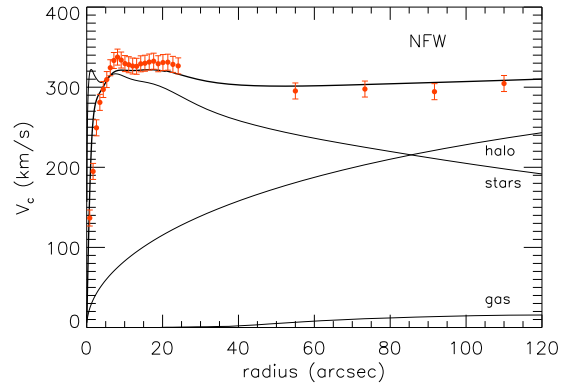,angle=0,width=8cm}}
\caption{Best fit model of a dark halo with an NFW profile, with a concentration parameter $c=10$ as indicated by cosmological simulation. This model has a stellar $M/L$ of 3.3 $M_\odot/L_{\odot,I}$. The red dots are the observations, and the black bold line the fit to these observations. The contributions of halo, stars and gas are plotted separately, where the first two curves are unconvolved. }
\label{fig:nfw_c10}
\end{figure}

Adopting $H_0 = 73$ km s$^{-1}$ Mpc$^{-1}$, the critical density is
given by $\rho_{\mathrm{crit}} = 3H_0^2/8 \pi G= 1.5\times10^{-7}$
$M_\odot$ pc$^{-3}$. We calculate the concentration parameter $c$,
given that

\begin{equation}
\frac{\rho_s}{\rho_{\mathrm{crit}}} = \frac{200}{3} 
 \frac{c^3}{\ln (1+c) - c/(1+c)},
\end{equation}

\noindent
and find $c=71$ and $c=4.7$ for the NFW profiles in the $M_*/L_I =$ 2.34
$M_\odot/L_{\odot,I}$ and $M_*/L_I =$ 3.8 $M_\odot/L_{\odot,I}$ models,
respectively. These values are quite deviant from the value that is
expected from cosmological simulations ($c \sim 10$, Bullock et al.
2001\nocite{2001MNRAS.321..559B}). When fixing $c=10$ and fitting
again an NFW halo to our observations with $M_*/L_I$ and the scale radius
as free parameters, we arrive at the model shown in
Figure~\ref{fig:nfw_c10}. We find $M_*/L_I =$ 3.3 $M_\odot/L_{\odot,I}$
and $r_s \approx 380^ {\prime \prime} \approx$ 38 kpc, with a
minimal $\chi^2$ value of 87 for $27-2$ degrees of freedom. We regard
this model as more realistic than the two other NFW profiles mentioned
above, but since also here the fit is not perfect, we cannot conclude
that therefore $M_*/L_I = $ 3.3 $M_\odot/L_{\odot,I}$ is a better
estimate for the stellar mass-to-light ratio in NGC~2974, than the value from the
stellar population models.

The results of the halo models discussed above are summarized in
Table~\ref{tab:haloes}.

\begin{table}
\begin{center}
\begin{tabular}{r|c|c|c|c|c}
\hline\hline
Halo profile & $M_*/L_I$  & $\rho_0$,$\rho_s$  & $r_c$,$r_s$  & $c$ & $\chi^2$ \\
 & ($M_\odot/L_{\odot,I}$) & ($M_\odot$ pc$^{-3}$) & (kpc) & & \\
\hline
Pseudo-  & 2.34 & 19 & 0.23 & - & 27 \\
isothermal  & 3.8 & 0.06 & 5.4 & - & 133 \\
\hline
 & 2.34 & 1.1 & 2.1 & 71 & 44 \\
NFW  & 3.8 & 0.0011 & 130 & 4.7 & 144  \\
 & 3.3 & 0.0067 & 38 & 10 & 87 \\ 
\hline
\end{tabular}
\end{center}
\caption{Comparison of the best fit models with a dark matter halo, as described in the text.}
\label{tab:haloes}
\end{table}

\subsection{MOND}

An alternative to including a dark matter halo in a galaxy to explain
its rotation curve at large radii, is provided by Modified Newtonian
Dynamics (MOND, Milgrom 1983\nocite{1983ApJ...270..365M}). In this
theory, Newtonian dynamics is no longer valid for small accelerations
($a \ll a_0$), but instead the acceleration $a$ in a gravitational
field is given by

\begin{equation}
\label{eq:mond}
a \mu(a/a_0) = a_N,
\end{equation}

\noindent
where $a_N$ is the Newtonian acceleration and $\mu$ is an
interpolation function, such that $\mu(x) = 1$ for $x \gg 1$ and
$\mu(x) = x$ for $x \ll 1$. Given the stellar mass-to-light ratio of a
galaxy, MOND predicts its rotation curve. An overview of properties
and predictions of MOND is offered by Sanders \& McGaugh
(2002)\nocite{2002ARA&A..40..263S}.

We fitted our rotation curve of NGC~2974 with $M_*/L_I$ as a free
parameter. For $a_0$ we adopted the value of $1.2 \times 10^{-8}$
cm/s$^2$, which was derived by Begeman, Broeils \& Sanders
(1991)\nocite{1991MNRAS.249..523B} from a sample of spiral galaxies.
The contribution of the neutral gas is included in our model in the
same way as described before, as well as a convolution to take seeing
and sampling into account.

NGC~2974 is an ideal candidate to study the transition between the
Newtonian and MOND regime, since the Newtonian acceleration reaches
$a_0$ at a radius of approximately $95^{\prime\prime}$ if we adopt a
stellar mass-to-light ratio of 2.34 $M_\odot/L_{\odot,I}$. For larger
$M_*/L_I$, this radius increases, and for the maximum disc value of 3.8
$M_\odot/L_{\odot,I}$, $a_0$ is reached around $120^{\prime\prime}$.
This means that a large part of the observed rotation curve lies in
the transition region, and we could therefore use NGC~2974 to
discriminate between interpolation function.

We first constructed a model with the standard interpolation function of MOND, 

\begin{equation}
\mu(x)= \frac{x}{\sqrt{1+x^2}}.
\end{equation}

\begin{figure}
    \centerline{\psfig{figure=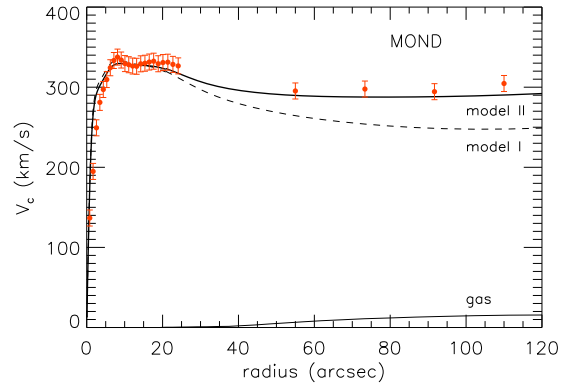,angle=0,width=8cm}}
\caption{Observed rotation curve (red datapoints) and best fitting models with a MOND rotation curve. The dashed line (Model I) is constructed with the standard interpolation function, while the thick line (Model II) uses an alternative simple interpolation function (see text).}
\label{fig:mond}
\end{figure}

\noindent
The resulting fit is shown as model I in Figure~\ref{fig:mond}. This
model has the same $M_*/L_I$ value as the maximal disc model, 3.8
$M_\odot/L_{\odot,I}$, but does clearly not provide a good fit to the
data. 

We constructed a second model, with an alternative interpolation
function explored by Famaey \& Binney (2005)\nocite{2005MNRAS.363..603F},

\begin{equation}
\mu(x) = \frac{x}{1+x}. 
\end{equation}

\noindent
This function makes the transition between the Newtonian and the MOND
region less abrupt than the standard interpolation function and
requires a lower $M_*/L_I$. The fit provided by this model to the data
is much better (Model II in Figure~\ref{fig:mond}), but formally
less good than a model with a dark matter halo. This model requires
$M_*/L_I = $3.6 $M_\odot/L_{\odot,I}$, and the fit yields $\chi^2 = 98$,
for 27-1 degrees of freedom.

Famaey \& Binney (2005)\nocite{2005MNRAS.363..603F} find that their
simple interpolation function provides better constraints to the
terminal velocity of the Milky Way than the standard function. Famaey
et al. (2007)\nocite{2007PhRvD..75f3002F} fitted the rotation curves
of a sample of galaxies with Hubble types ranging from small irregular
dwarf galaxies to large early-type spirals, and report that both
interpolating functions fit the data equally well. However, Sanders \&
Noordermeer (2007)\nocite{2007MNRAS.379..702S} find that for their
sample of early-type disc galaxies the simple interpolation function
yields more sensible values for $M_*/L$ than the standard one.

It would be interesting to see whether there is a preference for the
simple interpolation function over the standard one in early-type
galaxies. In this scenario, the challenge for MOND would be to provide
a universal interpolation function that would fit rotation curves of
all galaxy types along the Hubble sequence. So far, mostly spirals and
dwarf galaxies have been confronted with MOND, but with more
early-type galaxies getting detected in \HI\ and more rotation curves
becoming available, the sampling in morphology should become less
biased to late-type galaxies.

\section{Summary}
\label{sec:conclusion}

We obtained \HI\ observations of the early-type galaxy NGC~2974 and
found that the neutral gas resides in a ring. The ring starts around
50$^{\prime \prime}$, and extends to 120$^{\prime \prime}$, which
corresponds to 12 kpc or 5 effective radii. The total mass of the
neutral gas is $5.5 \times 10^8 M_\odot$.

We compared the velocity field of the \HI\ ring with the kinematics of
the ionised gas. We found that both velocity fields are very regular
and nicely aligned, indicating that they could form a single disc. A
harmonic decomposition of the velocity field showed that at large
radii the gravitational potential is consistent with an axisymmetric
shape.

We introduced a new way to correct the rotation curve of the ionised
gas for asymmetric drift. We found that the correction approaches a
constant value and this enabled us to remove the effect of turbulence
on the rotation curve, assuming a constant asymmetric drift correction
throughout the galaxy. We confirmed that this assumption is valid in
NGC~2974, by comparing with the asymmetric drift corrected rotation
curve of the stars (which was not affected by turbulence). An
interesting question is whether other galaxies show the same
behaviour. If this is the case, then with our method we would be able
to investigate rotation curves of ionised gas and the effects of
turbulence in more detail in other galaxies, and search for
connections with e.g. spiral structure and bars. Although in principle
we could for NGC~2974 also have used the stellar rotation curve
together with the \HI\ to constrain the mass models, this will not be
the case for all galaxies. For instance, in low surface brightness
galaxies stellar kinematics are not easy to obtain, and even in high
surface brightness galaxies, the absorption line kinematics need to be
binned to higher signal-to-noise than the emission line kinematics,
provided that ionised gas is present.

It is clear from the rotation curve of NGC~2974 that dark matter is
required to explain the observed velocities.We found that the total
mass-to-light ratio increases from 4.3 $M_\odot/L_{\odot,I}$ at 1
$R_e$ to 8.5 $M_\odot/L_{\odot,I}$ at 5 $R_e$. This last value would
correspond to 14 $M_\odot/L_{\odot,B}$ in $B$-band. Even in the maximal
disc model, 55 per cent of the total mass is dark, and an additional
dark halo needs to be included.

We constructed mass models of NGC~2974, where we modeled both the
stellar and gaseous contribution to the gravitational potential. The
latter is negligible compared to the stars ($M_{\mathrm{gas}} \sim
0.001 M_*$), but still included in our models. For the dark halo, we
tested two different profiles: the core-dominated pseudo-isothermal
sphere and the cuspy NFW profile. We experimented with different
values for the stellar mass-to-light ratio, but found that we cannot
constrain this value with just the rotation curve: for most $M_*/L$
smaller than the maximal disc value, we could obtain a decent fit for
both the pseudo-isothermal sphere and the NFW profile. If we compare
models with $M_*/L_I$ from single stellar population models (2.34
$M_\odot/L_{\odot,I}$) with maximal disc model ($M_*/L_I = $ 3.8
$M_\odot/L_{\odot,I}$), then the first provide better fits to the
data. Especially the inner datapoints are better reproduced with
$M_*/L_I =$ 2.34 $M_\odot/L_{\odot,I}$, but we note that the \HI\ data
points are better fitted in the models with the larger $M_*/L_I$ value.
The pseudo-isothermal sphere fits our data marginally better than the
NFW profile, but the difference is not significant. With MOND we can
also reproduce the observed rotation curve, but not as well as with
models that include a dark matter halo.

The largest uncertainty in our analysis is the stellar mass-to-light
ratio. We can only derive an upper limit on this ratio from the
maximal disc model or e.g. from Schwarzschild modeling, since $M_*/L$
should always be equal to or smaller than the dynamical $M/L$.  Values
for $M_*/L$ from stellar population synthesis models depend
significantly on the model assumptions: we mentioned already that
going from a Kroupa to a Salpeter IMF can increase $M_*/L$ by as
much as 40 per cent. Also, even low-level secondary star formation can
affect $M_*/L$ severely.  Furthermore, there is no reason why the
$M_*/L$ should remain constant over 5 effective radii, which we
assumed when modeling the stellar mass. If the stellar mass-to-light
ratio were known, we would be able to determine the dark matter
fraction in the galaxy with more accuracy, and either rule out or
confirm the maximal disc hypothesis. Also, since $M_*/L$ is the only
free parameter when fitting a rotation curve in MOND, knowing this
value would provide us with a rotation curve that can be compared to
the data directly, providing a clear test for MOND.

We have shown in this paper that it is possible to combine rotation
curves of neutral and ionised gas, correcting the latter one for
asymmetric drift using the Jeans equations and the higher order
velocity moments of the collisionless Boltzmann equations. Our method
to correct for the asymmetric drift therefore does not require a cold
disc assumption ($\sigma \ll V_c$). With more early-type galaxies
getting detected in \HI, and more high quality rotation curves
becoming available, we can now study the shape of their dark matter
haloes.


\section*{Acknowledgements}

The authors would like to thank Martin Bureau, Michele Cappellari,
Richard McDermid, Marc Sarzi and Scott Tremaine for useful
discussions, and Remco van den Bosch for careful reading of the
manuscript. We also are grateful to Jes\'us Falc\'on-Barroso for
making available the MDM image of NGC~2974 prior to publication.

This research was supported by the Netherlands Research School for
Astronomy NOVA, and by the Netherlands Organization of Scientific
Research (NWO) through grant 614.000.426 (to AW). AW acknowledges The
Leids Kerkhoven-Bosscha Fonds for contributing to working visits. GvdV
acknowledges support provided by NASA through Hubble Fellowship grant
HST-HF-01202.01-A awarded by the Space Telescope Science Institute,
which is operated by the Association of Universities for Research in
Astronomy, Inc., for NASA, under contract NAS 5-26555.

The Very Large Array is part of the National Radio Astronomy
Observatory, which is a facility of the National Science Foundation
operated under cooperative agreement by Associated Universities, Inc.
The \sauron\ observations were obtained at the William Herschel
Telescope, operated by the Isaac Newton Group in the Spanish
Observatorio del Roque de los Muchachos of the Instituto de
Astrof\'isica de Canarias.

The Digitized Sky Survey was produced at the Space Telescope Science
Institute under US Government grant NAG W-2166. Photometric data of
NGC~2974 was obtained using the 1.3-m McGraw-Hill Telescope of the MDM
Observatory at Kitt Peak. We acknowledge the usage of the HyperLeda
data base (http://leda.univ-lyon1.fr).



\appendix

\section{Asymmetric drift correction in a thin disc}
\label{sec:adc}

In this appendix we derive expressions for the asymmetric drift
correction in a stationary axisymmetric system, using the velocity
moments of the collisionless Boltzmann equation. We then evaluate this
expression in a thin disc approximation. Our method does not require
that the velocity dispersion should be small compared to the circular
velocity ($\sigma/V_c \ll 1$) and is comparable to the ``hot disc
model'', (see e.g. H\"aring-Neumayer et al.
2006\nocite{2006ApJ...643..226H}).

\subsection{The velocity ellipsoid}

To derive the asymmetric drift correction we start from the
collisionless Boltzman equation for a stationary axisymmetric galaxy
and using cylindrical coordinates $\vec r = (R, \phi, z)$,

\begin{equation}
\label{eq:cbe}
v_R\frac{\partial f}{\partial R} + v_z\frac{\partial f}{\partial z} + 
\Big( \frac{v_\phi^2}{R} - \frac{\partial \Phi}{\partial R}\Big) 
\frac{\partial f}{\partial v_R} - \frac{v_R v_\phi}{R} \frac{\partial f}
{\partial v_\phi} - \frac{\partial \Phi}{\partial z}
\frac{ \partial f}{\partial v_z} = 0,
\end{equation}

\noindent
with $f(R,z;v_R, v_\phi, v_z)$ the distribution function, $\Phi (R,
z)$ the underlying potential and $\nu(\vec r)$ the (luminosity) density
given by $\int f(\vec r; \vec v) d \vec v $. We multiply the above
equation by $v_R$ and subsequently integrate over all velocities. We
then obtain the Jeans equation:

\begin{equation}
\label{eq:jeans1}
\frac{\partial (\nu \overline{ v_R^2})} {\partial R} + 
\frac{\partial (\nu \overline{v_R v_z})}{\partial z} + \frac{\nu}{R}
\Big( \overline{v_R^2} - \overline{v_\phi^2} + R \frac{\partial \Phi}
{\partial R} \Big) = 0.
\end{equation}

\noindent
Since our system is axisymmetric, we set $\partial \nu / \partial
z = 0$ by symmetry. Substituting the circular velocity $V_c^2 =
R(\partial \Phi / \partial R)$, we arrive at Equation (4-33) of Binney
\& Tremaine (1987)\nocite{1987BT}:

\begin{equation}
\label{eq:adc1}
V_c^2 = \overline{v_\phi}^2 - \sigma_R^2 \Big[ \frac{\partial \ln \nu}
{\partial \ln R} + \frac {\partial \ln \sigma_R^2}{\partial \ln R} + 1 - 
\frac{\sigma_\phi^2}{\sigma_R^2} + \frac{R}{\sigma_R^2} 
\frac{\partial (\overline{v_R v_z})}{\partial z} \Big],
\end{equation}

\noindent 
with $\sigma_\phi^2 = \overline{v_\phi^2} - \overline{v_\phi}^2$,
$\sigma_R^2 = \overline{v_R^2}$ and $\sigma_z^2 =
\overline{v_z^2}$. The observed velocity field gives $\overline
{v_\phi}$, and the remaining terms in Equation~(\ref{eq:adc1}) form the
asymmetric drift correction.

The last term in the asymmetric drift correction depends on the
alignment of the velocity ellipsoid. In case of alignment with the
cylindrical coordinate system $(R, \phi, z)$ we have $\overline{v_R
  v_z} = 0$, while in case of alignment with the spherical coordinate
system $(r, \theta, \phi)$ we have $\overline{v_R v_z} = (\sigma_R^2 -
\sigma_z^2)(z/R)/[1 - (z/R)^2]$, which becomes proportional to $z/R$
close to the disc plane. These are two extreme situations, and we
introduce the parameter $\kappa$ to find a compromise:

\begin{equation}
\label{eq:kappa}
\overline{v_R v_z} = \kappa (\sigma_R^2 - \sigma_z^2)\frac{z/R}{1 - (z/R)^2},
 \qquad 0 \le \kappa \le 1,
\end{equation}

\noindent
where a typical value for $\kappa$ is 0.5 for disc galaxies (e.g. Kent
\& de Zeeuw 1991\nocite{1991AJ....102.1994K}).
 
To evaluate the asymmetric drift correction, we need expressions for
$\sigma_\phi / \sigma_R$ and $\sigma_z / \sigma_R$. We use higher
order velocity moments of the collisionless Boltzmann equation to
derive these expressions.

Starting again from Equation~(\ref{eq:cbe}), we multiply by $v_R v_\phi$
and integrate over all velocities:

\begin{equation}
\label{eq:jeans2}
\frac{\partial(\nu \overline{v_R^2 v_\phi})}{\partial R} + 
\frac{\partial(\nu \overline{v_R v_z v_\phi})} {\partial z} + \frac{\nu}{R}
\Big( 2\overline v_R^2 v_\phi - \overline{v_\phi^3} - \overline{v_\phi}
R\frac{\partial \Phi}{\partial R} \Big) = 0.
\end{equation}

\noindent
Aligning the velocity ellipsoid in the azimuthal direction we have
$\overline{v_R^2(v_\phi - \overline{v_\phi})} = 0$ and $\overline{v_R
v_z(v_\phi - \overline{v_\phi})} = 0$, so that $\overline{v_R^2
v_\phi} = \sigma_R^2\overline{v_\phi}$ and $\overline{v_R v_z v_\phi}
= \overline{v_R v_z}$ $\overline{v_\phi}$. We substitute these
relations in Equation~(\ref{eq:jeans2}), and subtract
$\overline{v_\phi}$ times the Jeans equation~(\ref{eq:jeans1}):

\begin{equation}
\nu \sigma_R^2 \frac{\partial \overline{v_\phi}}{\partial R} + \nu 
\overline{v_r v_z}\frac{\partial \overline{v_\phi}}{\partial z} + \frac{\nu}
{R} \Big[ \sigma_R^2 \overline{v_\phi} - \big(\overline{v_\phi^3} - \overline{v_\phi^2} \overline{v_\phi}\big) \Big] = 0.
\end{equation}

\noindent
We substitute $\overline{v_\phi^3} - \overline{v_\phi^2}
\overline{v_\phi} = 2 \sigma_\phi^2 \overline{v_\phi} + \overline{
(v_\phi - \overline{v_\phi}^3)}$ and $\overline{v_R v_z}$ from Equation(\ref{eq:kappa}) to arrive at

\begin{equation}
\label{eq:sigphisigr}
\frac{\sigma_\phi^2}{\sigma_R^2} = \frac{1}{2}\Big(1 + \alpha_R + \kappa
\frac{1 - \sigma_z^2/\sigma_R^2}{1 - (z/R)^2}\alpha_z
 - \frac{ \overline{(v_\phi - 
\overline{v_\phi})^3}} {\sigma_R^2 \overline{v_\phi}}\Big ), 
\end{equation}

\noindent
where we have introduced the logarithmic slopes

\begin{equation}
\alpha_R = \frac{\partial \ln \overline{v_\phi}}{\partial \ln R}, \quad
\textrm{and} \quad \alpha_z = \frac{\partial \ln \overline{v_\phi}}
{\partial \ln z}.
\end{equation}

To obtain an expression for $\sigma_z / \sigma_R$ we again start with
the collisionless Boltzmann equation, but now multiply with
$v_z(v_\phi - \overline{v_\phi})$ before integrating over all
velocities:

\begin{equation}
\nu \overline{v_R v_z} \frac{z}{R} \Big( 1+\frac{\partial \ln \overline{v_\phi}}
{\partial \ln R} \Big) + \nu \sigma_z^2\frac{\partial \ln \overline{v_\phi}}
{\partial \ln z} = 0.
\end{equation}

\noindent
Substituting Equation~(\ref{eq:kappa}) we find

\begin{equation}
\label{eq:sigzsigr}
\frac{\sigma_z^2}{\sigma_R^2} = \frac{\kappa z^2(1 + \alpha_R)}
{\kappa z^2(1 + \alpha_R) - (R^2-z^2)\alpha_z},
\end{equation}

The above expressions can be inserted into Equation~(\ref{eq:adc1}) to
obtain the asymmetric drift correction and therefore the true circular
velocity. In practice, we often apply the asymmetric drift correction
in the thin disc approximation, because from observations the
$z$-dependence is not straigthforward to derive. 

In the thin disc approximation, we have $z \ll R$, and therefore we
can write Equation~(\ref{eq:kappa}) as

\begin{equation}
\overline{v_Rv_z} = \kappa (\sigma_R^2 - \sigma_z^2) \frac{z}{R}.
\end{equation}

\noindent
and following the same reasoning as before, we see that the
expressions in Equations~(\ref{eq:sigphisigr}) and (\ref{eq:sigzsigr}) simplify
slightly: in the first expression the one-to-last term disappears, and for
the second one, $(R^2-z^2)$ gets replaced by $R^2$ in the nominator of the
expression.  Furthermore, the derivative of $\overline{v_Rv_z}$
simplifies considerably.

We find the following expression for the asymmetric drift
correction in the thin disc approximation, after substitution in
Equation~(\ref{eq:adc1}):

\begin{eqnarray}
\label{eq:adc2}
V_c^2 = \overline{v_\phi}^2 -  \sigma_R^2 \Big[ \frac{\partial \ln \nu}
{\partial \ln R} + \frac{\partial \ln \sigma_R^2}{\partial \ln R} + 
\frac{1}{2}(1-\alpha_R) + \nonumber\\ 
 \frac{1}{2}\frac{\overline{(v_\phi - \overline{v_\phi})^3}}
{\sigma_R^2 \overline{v_\phi}} - \frac{\kappa R^2 \alpha_z}
{\kappa z^2 (1 + \alpha_R) - R^2\alpha_z}\Big].
\end{eqnarray}

\noindent
The one-to-last term vanishes in the case of a velocity ellipsoid
symmetric around $v_\phi = \overline{v_\phi}$. This need not
necessarily be the case, and the exact form of $\overline{(v_\phi -
  \overline{v_\phi})^3}$ depends on the underlying distribution
function, which in general cannot be constrained easily (e.g. Kuijken
\& Tremaine 1991\nocite{1991dodg.conf...71K}). However, since this
term is a factor $\sigma_R^2$ smaller than the other terms, it can be
safely ignored for most purposes.

\subsection{Observables}
\label{sec:adc_obs}

Here we investigate how in the thin disc approximation we can correct
our observed velocity field for asymmetric drift, to obtain the
true circular velocity $V_c$. This quantity traces the potential and
therefore the mass of the galaxy.

In a thin disc, we can replace $\partial \ln \nu / \partial \ln R$ by
the slope of the surface brightness $\partial \ln \Sigma / \partial
\ln R$. This slope can be obtained directly
from observations. 

The observed velocity and velocity dispersion of an axisymmetric thin
disc seen under an inclination $i$ is given by:

\begin{eqnarray}
\label{eq:obs}
V & = &v_{\mathrm{sys}} + \overline{v_\phi}\cos \phi \sin i, \nonumber\\
\sigma^2 & = & \sigma_R^2 \sin^2\phi \sin ^2 i + \sigma_\phi^2 \cos^2\phi
 \sin^2 i +  \sigma_z^2\cos^2 i \nonumber\\
 & & - \overline{v_R v_z} \sin \phi \sin{2i}.
\end{eqnarray}

\noindent
It is straightforward to obtain $\overline{v_\phi}$ from the observed
velocity field, and though $\alpha_R$ can be estimated rather well,
$\alpha_z$ is less easy to constrain. Therefore, we fit to
$\overline{v_\phi}$ the prescription of Evans \& de Zeeuw
(1994\nocite{1994MNRAS.271..202E}) for power-law models:

\begin{equation}
\label{eq:powerlaw}
v_{\mathrm{mod}}\propto \frac{R}{(R_c^2+R^2 + z^2/q_\Phi^2)^{1/2 + \beta/4}},
\end{equation}

\noindent
where $R_c$ is the core radius, $q_\Phi$ the flattening of the
potential and $\beta$ the logarithmic slope of the rotation curve at
large radii (such that $\beta = 0$ implies a flat rotation curve).

For the slopes of $\overline{v_\phi}$ we find that:

\begin{eqnarray}
\label{eq:alpha}
\alpha_R & = &1 - \frac{(1+\beta/2)R^2}{R_c^2+R^2+z^2/q_\Phi^2} \nonumber\\
\alpha_z & = &- \frac{(1+\beta/2)z^2/q_\Phi^2}{R_c^2+R^2+z^2/q_\Phi^2} = 
 - \frac{z^2}{q_\Phi^2 R^2} (1 - \alpha_R),
\end{eqnarray}

\noindent
so that with $\overline{v_R v_z} = 0$ in the disc plane, we obtain:

\begin{eqnarray}
\sigma^2 & = & \sigma_R^2 \big[ 1 - \frac{1}{2}(1\!-\!\alpha_R)\cos^2\phi \sin^2 i - 
\nonumber\\ & & \frac{(1+\beta/2)R^2}{\kappa q_\Phi^2 (R_c^2\!+\!R^2)(1\!+\!\alpha_R) + 
(1\!+\!\beta/2)R^2} \cos^2 i \big].  
\end{eqnarray}

\noindent
When evaluated along the major axis, $\cos^2\phi=1$. 

Assuming that the velocity ellipsoid is symmetric around $v_\phi =
\overline{v_\phi}$ the corresponding term in Equation~(\ref{eq:adc2})
vanishes. Inserting the relations obtained from the power-law model,
we arrive at the following expression for the circular velocity:

\begin{eqnarray}
\label{eq:adc3}
V_c^2 = \overline{v_\phi}^2 - \sigma_R^2 \Big[\frac{\partial \ln \Sigma}
{\partial \ln R} + \frac{\partial \ln \sigma_R^2}{\partial \ln R} + 
\frac{1}{2} (1-\alpha_R) + \nonumber\\ \frac{\kappa(1 - \alpha_R)}
{\kappa(2R_c^2-R^2) + R^2}\Big].
\end{eqnarray}

\label{lastpage}
\end{document}